\documentclass[twocolumn]{aastex7}
\usepackage{caption}

\newcommand{\Heii}{\textrm{He\,\textsc{ii}}}
\newcommand{\Feii}{\textrm{Fe\,\textsc{ii}}}
\newcommand{\Mgii}{\textrm{Mg\,\textsc{ii}}}
\newcommand{\Civ}{\textrm{C\,\textsc{iv}}}
\newcommand{\Oiii}{[\textrm{O\,\textsc{iii}}]}
\newcommand{\Ciii}{\textrm{C\,\textsc{iii}}]}

\begin{document}

\title{A Sample of He\,\textsc{II} $\lambda$4686 ``Changing-Look" Quasars}

\author{Wen-Tao Lu}
\affiliation{\rm Department of Astronomy, University of Science and Technology of China, Hefei, Anhui, 230026, PR China}
\email[show]{loneica@mail.ustc.edu.cn}  

\author{Jun-Xian Wang}
\affiliation{\rm Department of Astronomy, University of Science and Technology of China, Hefei, Anhui, 230026, PR China}
\affiliation{\rm School of Astronomy and Space Science, University of Science and Technology of China, Hefei, Anhui 230026, PR China} 
\affiliation{\rm College of Physics, Guizhou University, Guiyang, Guizhou, 550025, PR China}

\email[show]{jxw@ustc.edu.cn}

\begin{abstract}

We present the first systematic search for ``changing-look" (``CL") behavior in the broad \Heii\ $\lambda$4686 emission line in quasars, utilizing repeated spectroscopy from the Sloan Digital Sky Survey (SDSS). The \Heii\ line, originating from high-ionization gas and powered by extreme ultraviolet photons, serves as a sensitive tracer of changes in the ionizing continuum. After applying strict spectral selection criteria and visual inspection to a parent sample of over 9,000 quasars with multi-epoch spectra, we identify a sample of 34 \Heii\ “changing-look” quasars that show a significant appearance or disappearance of the broad \Heii\ $\lambda$4686 line. Compared with previously known H$\beta$ “CL” quasars, the \Heii\ “CL” sample exhibits similarly strong continuum variability and broad-line flux changes, yet shows a preference for higher Eddington ratios and lower host-galaxy contamination. These results highlight the value of \Heii\ line in studying the central variable engines of AGNs and uncovering a more complete census of extreme quasar variability. A comparison with H$\beta$ “CL” further underscores the profound selection biases inherent in “changing-look” studies, especially those associated with line strength, host-galaxy contamination, and spectral signal-to-noise ratio.

\end{abstract}


\section{Introduction} 

Active Galactic Nuclei (AGNs) are typically divided into two spectral types — type 1 and type 2 — depending primarily on the presence (type 1) or absence (type 2) of broad emission lines (BELs) in their optical spectra. Intermediate subtypes (1.2, 1.5, 1.8, 1.9) are defined by progressively weaker broad Balmer lines, particularly H$\beta$ and H$\alpha$ \citep[e.g.,][]{1976MNRAS.176P..61O,1977ApJ...215..733O,1981ApJ...249..462O}. A widely used quantitative criterion for spectral classification relies on the flux ratio between the broad H$\beta$ and narrow \Oiii\ $\lambda5007$ lines, defined as $R = F_{\rm H\beta}/F_{\rm [OIII]}$: type 1.0 AGNs have $R > 5$; type 1.2 span $2 < R < 5$; type 1.5 have $0.33 < R < 2$; type 1.8 have $R < 1/3$; type 1.9 AGNs exhibit only a broad H$\alpha$ component without detectable broad H$\beta$, and type 2.0 AGNs show no broad Balmer lines \citep{10.1093/mnras/257.4.677}.  In practice, subtypes 1.0 through 1.5 are often grouped as type 1, while subtypes 1.8, 1.9, and 2.0 are collectively referred to as type 2 AGNs. Within the AGN unification framework, this dichotomy is attributed primarily to orientation effects, where an obscuring dusty torus blocks the central engine and the broad-line region from view in type 2 sources \citep[e.g.][]{1993ARA&A..31..473A}.

Since the 1970s, a rare subclass of AGNs has been discovered, the so-called changing-look AGNs (CLAGNs), which exhibit a transition between type 1 and type 2 spectral classifications (or vice versa) on timescales of months to years, due to the emergence or disappearance of broad Balmer emission lines. Several early examples include Mrk 590 (NGC 863), which evolved from a Seyfert 1.5 in the 1970s \citep{1977ApJ...215..733O} to a classical type 1 in the 1980s \citep[e.g.,][]{1990ApJ...363L..21F}, but subsequently lost its BELs and became a type 2 AGN by the 2010s \citep[e.g.,][]{2014ApJ...796..134D}. Another well-studied case is Mrk 1018, which transitioned from type 1.9 to type 1 between 1978 and 1984 \citep{1986ApJ...311..135C}. Even more dramatically,  \citet{10.1093/mnrasl/slt154} reported a CLAGN that changed from type 2 to type 1 within just 20 days.   

In the past decade, large-scale spectroscopic surveys have also uncovered analogous phenomena in luminous AGNs, or changing-look quasars (CLQs), prompting intense interest \citep[e.g.,][]{2015ApJ...800..144L,2016MNRAS.457..389M,Sheng2017,2018ApJ...862..109Y}.

Two major physical mechanisms have been proposed to explain the changing-look phenomenon: variable obscuration by intervening material (e.g., the torus) and intrinsic changes in the central accretion activity. 
Though X-ray monitoring has indeed revealed large changes in absorbing column densities in some AGNs \citep[e.g.,][]{2003MNRAS.342..422M,2002ApJ...571..234R,Miniutti_2013}, the majority of spectral transitions in CLAGNs are best interpreted as arising from intrinsic changes in accretion \citep[e.g.,][]{Sheng2017,2016MNRAS.457..389M}. 

Notably, in early studies, CLAGNs were defined by transitions between type 1 and type 2 spectra—namely, the appearance or disappearance of broad Balmer lines. In dim states, these sources may appear as type 1.8, 1.9, or even type 2 AGNs. More recently, as studies have expanded to higher redshifts, where lines such as \Mgii, \Civ, \Ciii, and Ly$\alpha$ are accessible from the ground, the term “changing-look” is often used more generally to denote the emergence or disappearance of a specific broad line, which does not necessarily correspond to a traditional type 1/2 spectral transition. Hereafter in this paper, we adopt the term ``changing-look” (or ``CL") in this generalized sense.

These recent findings reveal that different broad emission lines do not always vary synchronously during ``changing-look" events. For example, \citet{Guo_2019} reported a case where broad \Mgii\ reappeared while broad H$\beta$ and H$\alpha$ remained undetected, implying a spectral change not captured by classical Balmer-line classification. Similarly, \citet{2018ApJ...862..109Y} found an AGN (J1104+6343) with prominent variability in \Mgii\ but little change in H$\beta$. Moreover, it is often observed that in CLAGNs selected based on broad H$\beta$ variability, the broad \Mgii\ line remains visible even in the dim state \citep[e.g.][]{Guo_2019,2016MNRAS.457..389M,10.1093/mnras/staa645}. 
In the UV, \citet{2020MNRAS.498.2339R} presented several \Civ\ CLAGNs, including one with negligible variation in Ly$\alpha$, while \citet{guo2024identificationlymanalphachanginglook} reported a Ly$\alpha$ CLAGN with steady \Civ\ emission. This behavior might be attributed to a non-reverberating core component of \Civ\  driven by outflows \citep{Denney_2012,Wang_2020}.
Furthermore, \citet{2024ApJS..270...26G} conducted a systematic search for ``changing-look" events in H$\alpha$, H$\beta$, \Mgii, \Ciii, and \Civ\ using DESI data and revealed complex relationships among different broad emission lines.
Together, these studies demonstrate that ``changing-look" behavior is not necessarily synchronized across different broad emission lines, and highlight the importance of using multiple emission lines to diagnose the physical nature of variable central engines in AGNs.

In this work, we investigate the ``changing-look" behavior of the broad \Heii\ $\lambda$4686 line, a high-ionization feature with an ionization potential four times that of hydrogen lines. Unlike low-ionization lines such as the Balmer series and \Mgii, \Heii\ $\lambda$4686 is sensitive to much shorter-wavelength (extreme-UV) ionizing photons, offering a more direct probe of continuum variations at these energies. This makes it a distinctive diagnostic of the physical drivers behind ``changing-look“ events in quasars.

Dramatic changes in broad \Heii\ $\lambda$4686 emission have been observed in several individual AGNs. For instance, in 3C 390.3, the \Heii\ line nearly disappeared for a period while H$\beta$ remained stable \citep{10.1093/mnras/staa1210}. In NGC 4051, both the X-ray continuum and the \Heii\ line were observed to drop significantly over time, suggesting a transition to an advection-dominated accretion state \citep{Peterson_2000}. Although \citet{MacLeod_2019} primarily searched for H$\beta$ CLAGNs, they reported three quasars with extreme \Heii\ variability—only one of which showed concurrent changes in H$\beta$. 

Despite these intriguing cases, a systematic search for \Heii\ ``changing-look" quasars has yet to be performed. In this work, we conduct such a search using repeated SDSS spectroscopy and identify 34 \Heii\ ``changing-look" quasars.

The remainder of this paper is organized as follows. In \S\ref{sec:data}, we introduce the parent quasar sample, detail our spectral fitting methodology, and outline the criteria used to identify \Heii\ “changing-look” quasars. In \S\ref{sec:results}, we present the resulting sample and its key properties. In \S\ref{sec:discussion}, we explore the nature of these \Heii\ “changing-look” quasars through comparisons with previously known H$\beta$ counterparts. Finally, \S\ref{sec:summary} summarizes our main conclusions.

Throughout this work, we assume a flat $\Lambda$CDM cosmology with parameters $\Omega_{\Lambda} = 0.7$, $\Omega_\mathrm{m} = 0.3$, and a Hubble constant of $H_0 = 70\ \mathrm{km\ s^{-1}\ Mpc^{-1}}$.

\section{Data, Spectral Fitting and Selection of \Heii\ ``Changing-look" Quasars} \label{sec:data}

\subsection{Data and the parent sample}\label{sec:parentsample}
To select \Heii\ ``changing-look" quasars, we utilize quasars or galaxies with multiple spectroscopic observations from SDSS DR17 \citep{Abdurrouf_2022}, identified by their bestObjID value, and require at least one of the spectroscopic observations is classified as an quasar. 
Since we would like to compare the variation of \Heii\ line with that of H$\beta$, we restrict the sample to sources with redshifts $z < 0.8$, ensuring that the H$\beta$ line remains within reliable spectral coverage.
To exclude mismatched or erroneous associations, we further require that in at least half of the spectra, the redshift difference between any two epochs must not exceed 0.01. The quality of each spectrum is ensured by requiring a median signal-to-noise ratio (S/N) greater than 3 across all pixels in the spectrum.

To identify \Heii\ ``changing-look" events, we perform pairwise comparisons of all available spectra for each source. For objects with dense spectroscopic sampling, such as those from the SDSS Reverberation Mapping (SDSS-RM) program \citep{Shen2015}, this would result in too large number of spectral pairs. Therefore, SDSS-RM sources are excluded from this general selection and are analyzed separately (see \S\ref{sec:RMselection}). 

After applying the above criteria, we obtain a sample of 9,892 quasars with 23,964 spectroscopic pairs. We further exclude spectra that show large deviations in \Oiii\ $\lambda$5007 line flux between epochs, which are indicative of severe flux calibration issues under the assumption that this narrow emission line remains intrinsically constant (see \S\ref{sec:specfitting} for details). This results in a final parent sample of 9,363 quasars and 21,924 spectral pairs.

To characterize the photometric variability of the identified \Heii\ ``CL" quasars,  incorporate photometric data from SDSS, Pan-STARRS \citep{chambers2019panstarrs1surveys,Magnier_2020,Waters_2020,Flewelling_2020}, and ZTF \citep{Masci_2019,Bellm_2019,https://doi.org/10.26131/irsa598}.All the {\it Pan-STARRS} data used in this paper can be found in MAST:\citep{https://doi.org/10.17909/s0zg-jx37}.
 For SDSS and Pan-STARRS, we use the PSF flux (psfFlux), which is consistent with the spectroscopic SPECTROFLUX for point-like sources. For ZTF, we use the reported magnitudes (mag) without additional correction, given ZTF’s lower sensitivity and significantly coarser spatial resolution compared to SDSS and Pan-STARRS.

\subsection{Spectral fitting}\label{sec:specfitting}

We perform spectral fitting for all available spectra using the PyQSOFit package \citep{2018ascl.soft09008G,2019ApJS..241...34S}.

To account for host galaxy contamination, we first apply a Bayesian PCA-based host subtraction technique \citep{2024ApJ...974..153R}\footnote{While performing the PyQSOFit decomposition, we followed the procedure of \cite{2024ApJ...974..153R}, adopting the first five galaxy eigenspectra from \citet{Yip_2004q} and the first ten quasar eigenspectra from \citet{Yip_2004g}. Note the quasar eigenspectra were derived from a subsample of luminous quasars chosen to minimize host-galaxy contamination, as emphasized in \cite{2024ApJ...974..153R}. On the other hand, the first five galaxy eigenspectra provided by \citet{Yip_2004g} exhibit negligible broad-line components, implying that QSO contributions, if any, are very limited. This combination helps reduce the risk of cross-contamination between AGN and stellar light in our spectral decomposition. Meanwhile, \cite{2024ApJ...974..153R} reported good agreement between this method and host–AGN decompositions from imaging, without clear evidence of systematic bias.}. For each source, we select the spectrum with the faintest synthetic $g$-band spectroscopic magnitude, corresponding to the epoch with the weakest AGN contribution, and perform AGN-host spectral decomposition on that spectrum. The resulting host component is then consistently subtracted from all available spectra of the same source.

Secondly, the remaining AGN continuum is modeled as a combination of a power-law component, a low-order polynomial, and an empirical \Feii\ pseudo-continuum, fitted over a set of relatively line-free spectral windows.

After subtracting the continuum, we fit the \Heii\ $\lambda$4686, H$\beta$, and \Oiii\ complex using a total of nine Gaussian components: two broad Gaussians for the H$\beta$ broad emission line, one Gaussian for the H$\beta$ narrow component, four Gaussians for \Oiii\ $\lambda$4959 and $\lambda$5007 (two for the core components and two for potential outflow-driven wing components), and two Gaussians for the \Heii\ $\lambda$4686 emission (one broad and one narrow). During the fitting, the widths and velocity offsets of all narrow components, including the \Oiii\ core lines, narrow H$\beta$, and narrow \Heii, are tied together. The flux ratio between \Oiii\ $\lambda$4959 and $\lambda$5007 are always constrained to 1:3. We set an upper limit of 15,000 km/s on the FWHM of broad components, as unusually broad features are often artifacts resulting from poorly modeled continua. Moreover, when a broad emission line is genuinely that wide, its identification becomes increasingly difficult through visual inspection.

To mitigate the impact of flux uncertainties, such as those caused by fiber drop/loss issues \citep[e.g.,][]{2013AJ....145...10D, Margala_2016}, we recalibrate the multi-epoch spectra of each source using the \Oiii\ $\lambda$5007 flux as a reference. For this specific purpose, we derive the \Oiii\ flux by fitting the \Oiii\ doublet with two Gaussians (instead of four) to avoid additional uncertainties arising from the decomposition into core and wing components.
We first discard all spectra with abnormally low or high \Oiii\ $\lambda$5007 fluxes, defined as those deviating by more than a factor of two from the average flux (i.e., outside the range $0.5\overline{F}$ to $2\overline{F}$). For the remaining spectra, we recalibrate the spectra according to their \Oiii\ flux deviation from the average flux: any spectrum with \Oiii\ $\lambda$5007 flux differing by more than 20\% (i.e., outside the range $0.8\overline{F}$ to $1.25\overline{F}$) is adjusted accordingly. For sources with recalibrated spectra, we repeat the full set of spectral analysis procedures, including AGN–host decomposition, host subtraction, AGN continuum modeling, and emission-line fitting.

\begin{figure}
\includegraphics[width=1.0\linewidth]{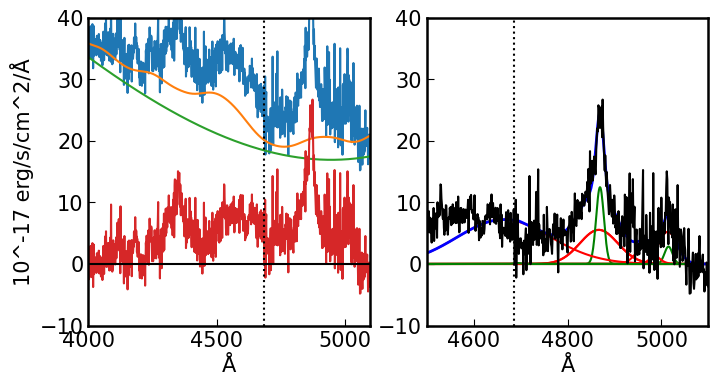}
\includegraphics[width=1.0\linewidth]{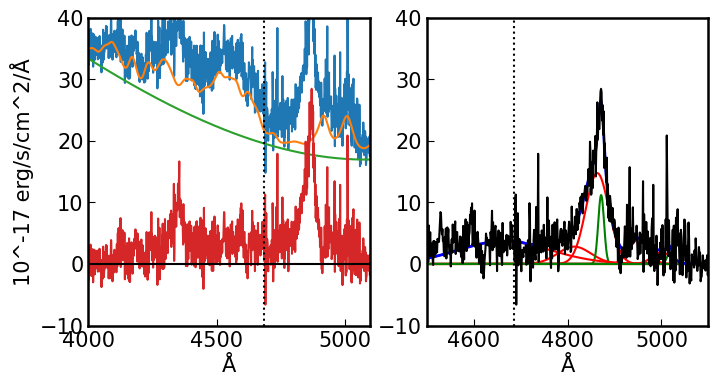}
\caption{Two example artificial spectra (upper and lower panels) generated by adding random Gaussian noise to a real SDSS quasar spectrum with strong \Feii\ emission (SDSS J090112.04+011549.5). In the left panels, the host-subtracted spectra are shown in blue, with the best-fit power-law continuum and power-law plus \Feii\ components over-plotted in green and yellow, respectively. The red lines indicate the residual spectra after subtracting the model continuum. Vertical dotted lines mark the position of the \Heii\ 4686 emission line. In the right panels, the residual spectra are re-plotted alongside the corresponding best-fit \Heii\ and H$\beta$ emission line models. This figure illustrates the degeneracy between \Feii\ and \Heii\ 4686, and highlights how inaccurate \Feii\ modeling can lead to spurious broad \Heii\ features in the residual spectrum.
\label{fig:Feii}}
\end{figure}

It is important to note that \Heii\ 4686 is heavily blended with the strongest part of the \Feii\ pseudo-continuum, introducing complex degeneracies and systematic uncertainties in spectral fitting, particularly in quasars with strong \Feii\ emission.
In Fig. \ref{fig:Feii}, we show two example artificial spectra generated by adding random Gaussian noise to a real SDSS quasar spectrum, along with their corresponding best-fit continuum and \Feii\ components. As illustrated, particularly in the upper panel of Fig. \ref{fig:Feii}, in cases where the \Feii\ component is not properly modeled, residuals can produce an artificial broad \Heii\ feature. These degeneracies and systematic uncertainties highlight the limitations of single-fit approaches, which we address through Monte Carlo simulations in the following sub-section.

\subsection{Selection of \Heii\ ``CL" events}\label{sec:selection}

We identify candidate \Heii\ ``changing-look" events based on the following criteria. For each pair of spectra, we designate the spectrum with the highest broad \Heii\ $\lambda$4686 flux as the bright state, and the one with the lower flux as the dim state. To ensure the line is visually prominent in the bright-state spectrum, we calculated the equivalent width (EW) of the broad \Heii\ $\lambda$4686 using the continua that includes the host galaxy contribution, and require the EW to exceed 10 \AA. Additionally, the \Heii\ flux in the bright state must be at least 2.5 times greater than in the dim state. Applying these criteria yields 3,027 galaxies with a total of 5,009 spectral pairs.

The next step involves visual inspection. Due to the strong degeneracy between the \Heii\ broad emission line and the underlying \Feii\ pseudo-continuum—as well as potential contamination from the nearby HeI $\lambda$4471 line—even visual assessment can be inconclusive in some cases. To ensure high sample purity, we require that the bright-state spectrum exhibits a visually prominent bump near 4686 \AA\ that is absent in the dim-state spectrum. Applying this visual selection yields 89 galaxies (89 spectral pairs) for further analysis.

To assess the significance of \Heii\ flux variability while accounting for the aforementioned degeneracy between \Feii\ and \Heii\ emission, we perform 500 Monte Carlo iterations for each spectrum. In each iteration, we add Gaussian noise to the observed spectrum based on its flux uncertainty at each wavelength, and then repeat the full spectral fitting procedure described in \S\ref{sec:specfitting}, including host decomposition and subtraction, AGN continuum and \Feii\ modeling, and emission line fitting. From the resulting distribution of fitted broad \Heii\ fluxes, we estimate the flux uncertainties using the 16th and 84th percentiles (i.e., the central 68\% interval).

We then quantify the significance of variability by requiring the flux difference between the bright and dim states to satisfy
$(F_{bright}-F_{dim})/\sqrt{\sigma_{bright}^2+\sigma_{dim}^2}$ $>$ 3, 
where the uncertainties $\sigma_\mathrm{bright}$ and $\sigma_\mathrm{dim}$ are derived from the Monte Carlo flux distributions. Finally, we visually inspect the candidate spectra to ensure both the robustness of the fits and the consistency of the flux distributions, resulting in a final sample of 31 \Heii\ ``CL" quasars.

\begin{figure*}
\hspace{1cm}
    \includegraphics[width=0.9\linewidth]{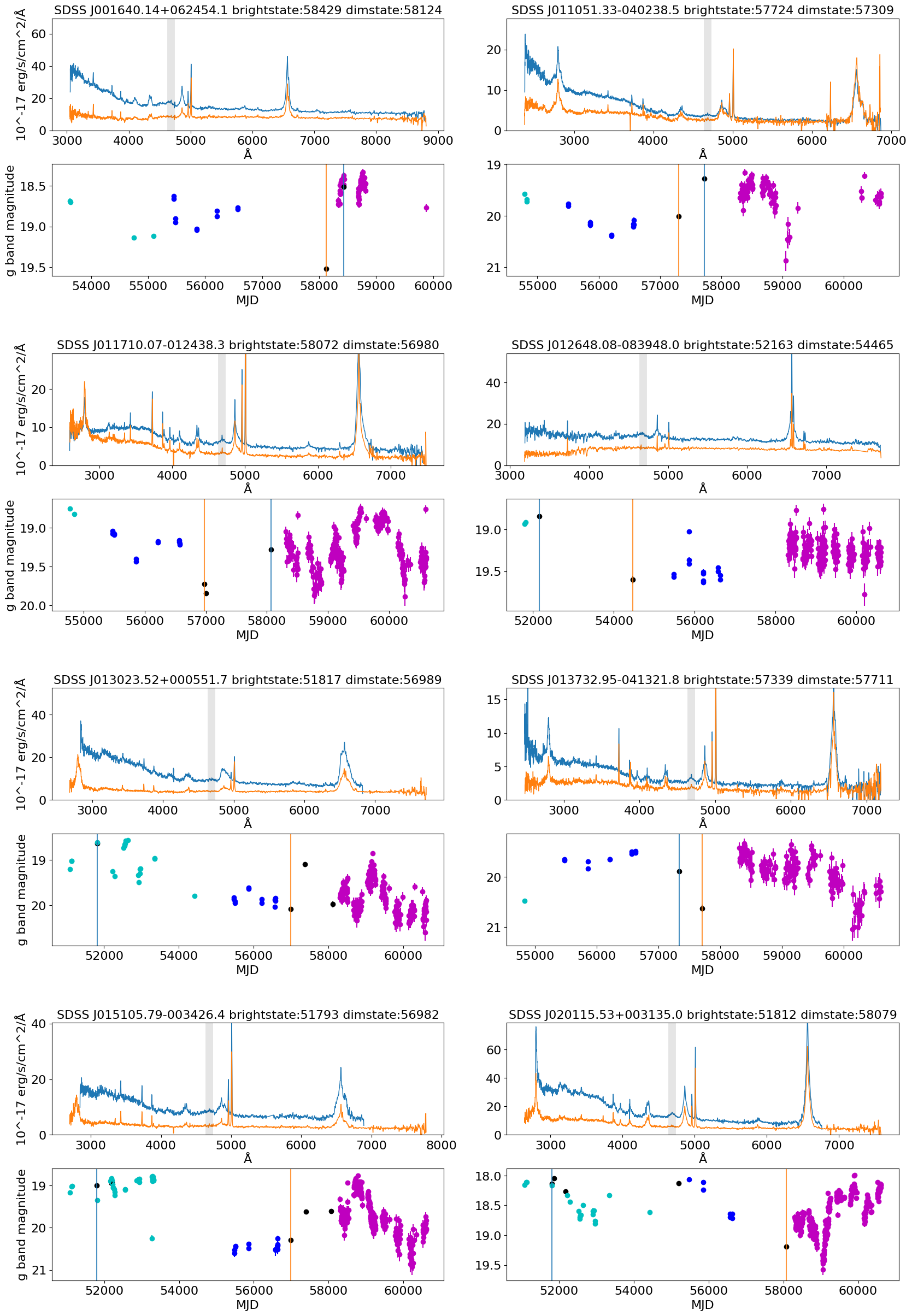}
    \caption{ }
    \label{fig:spectra1}
\end{figure*}
\begin{figure*}
    \includegraphics[width=0.9\linewidth]{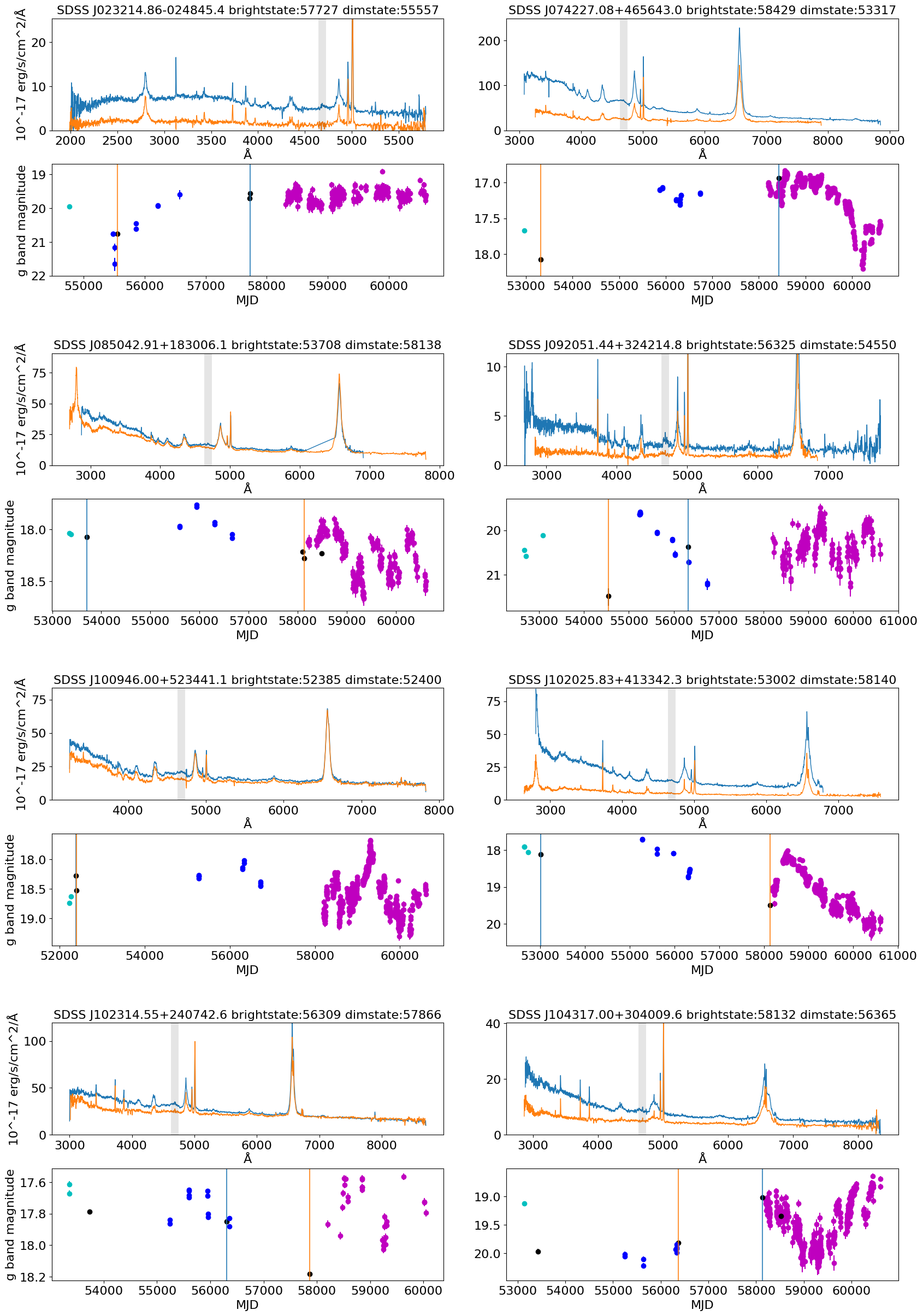}  
    \caption{ }
    \label{fig:spectra2}
\end{figure*}
\begin{figure*}
    \includegraphics[width=0.9\linewidth]{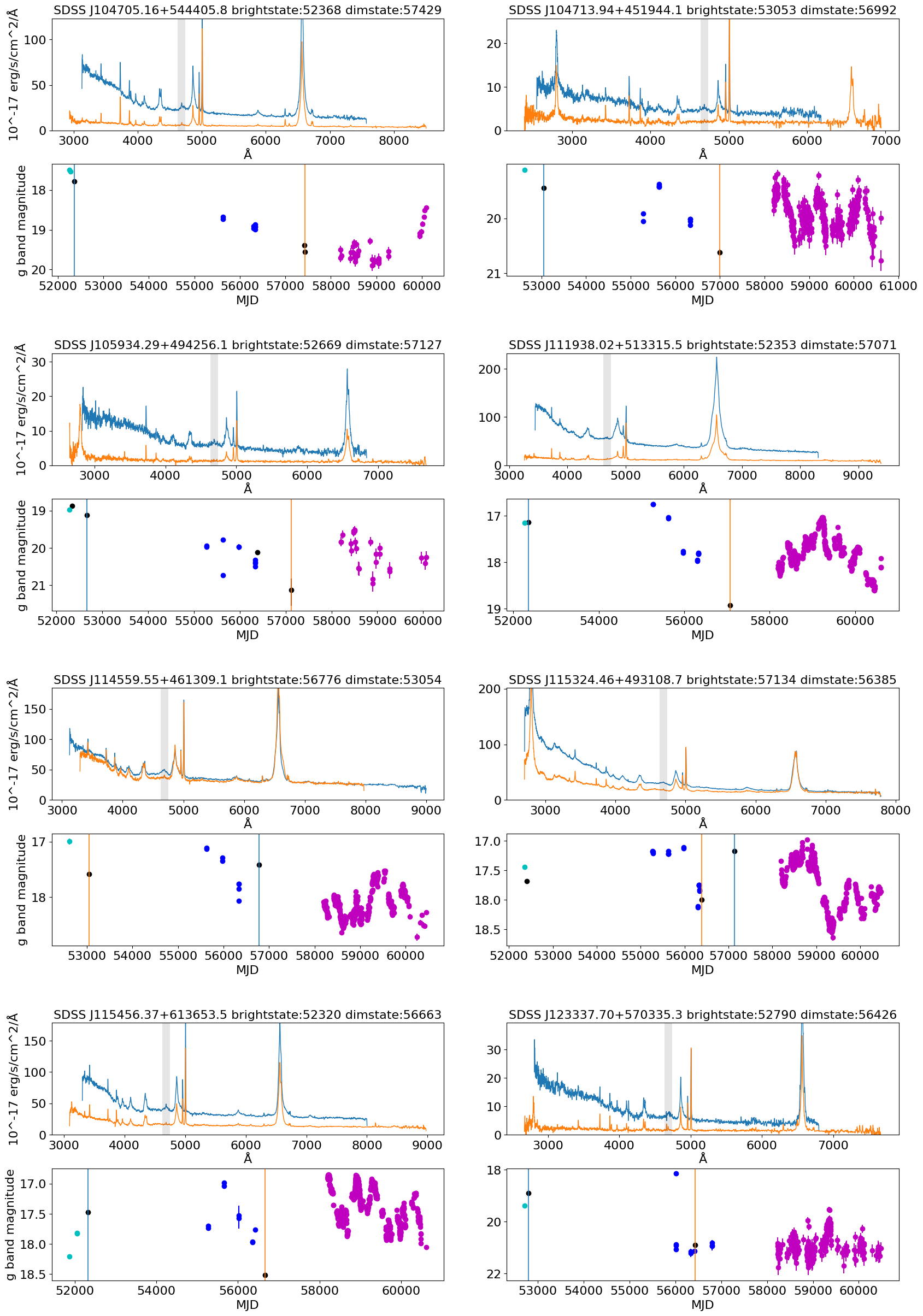}
    \caption{ }
   \label{fig:spectra3}
\end{figure*}
\begin{figure*}
\hspace{1cm}
    \includegraphics[width=0.9\linewidth]{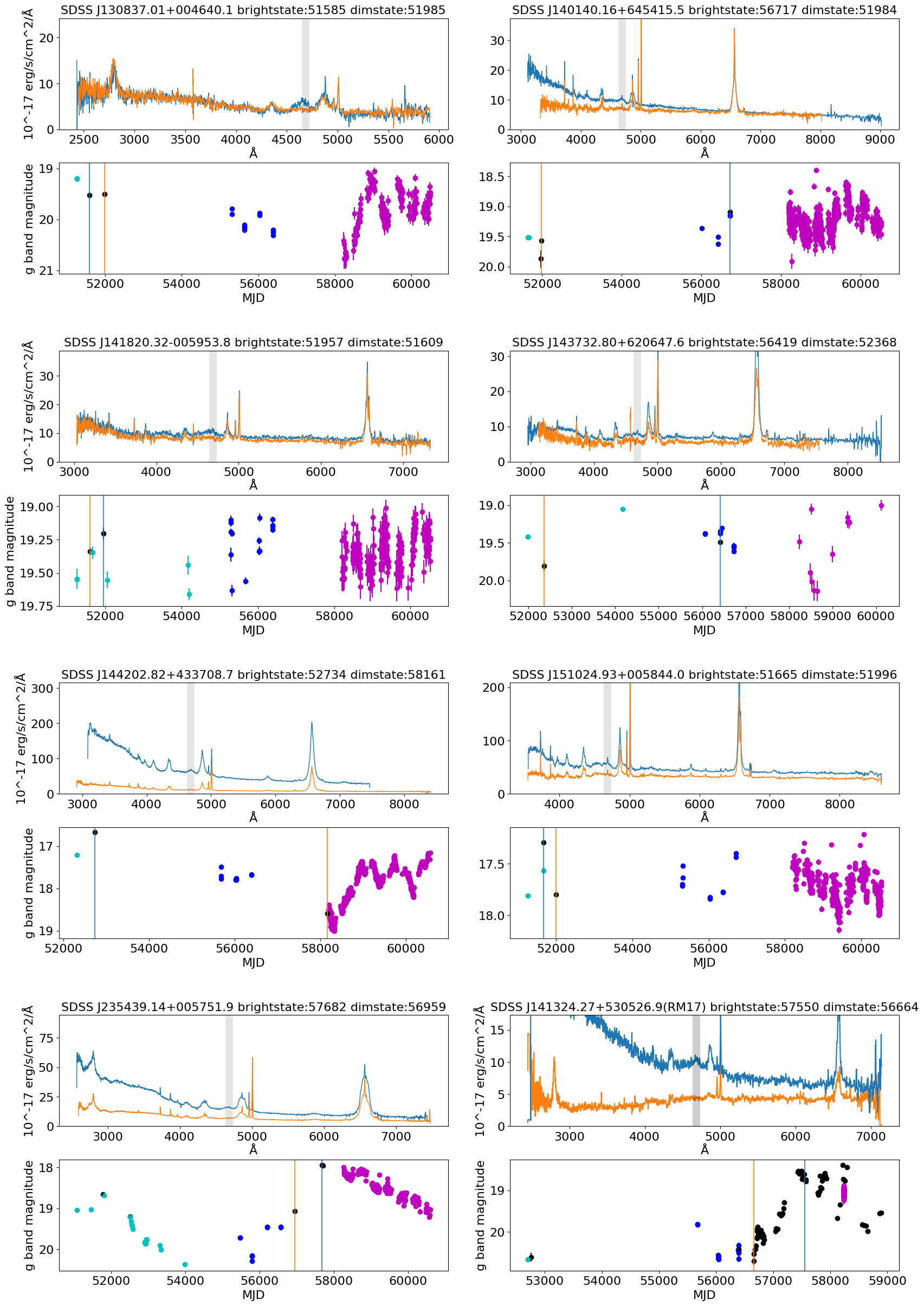}
    \caption{ }
    \label{fig:spectra4}
\end{figure*}
\begin{figure}
    \includegraphics[width=0.9\linewidth]{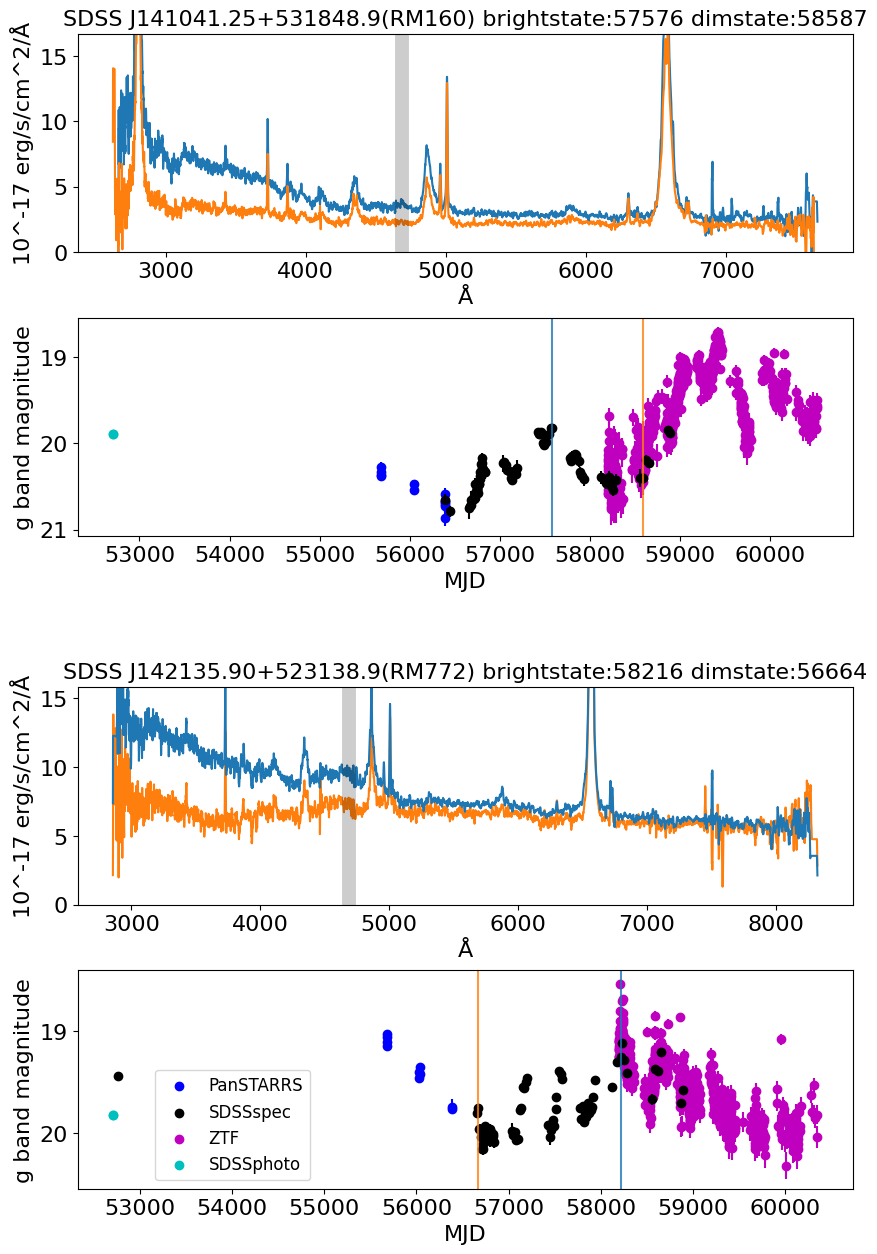}  
    \caption{
    In Figure \ref{fig:spectra1} to \ref{fig:spectra5} we show the bright-state (blue) and dim-state (orange) SDSS spectra for the 34 \Heii\ ``changing-look" quasars identified in \S\ref{sec:selection}. All spectra have been smoothed using a 5-pixel boxcar kernel. The shaded gray region marks the location of the \Heii\ $\lambda$4686 broad emission line. Each panel is labeled with the object name and the observation dates of the two spectral epochs. The corresponding $g$-band light curve is shown below each spectrum, combining SDSS photometry (cyan), synthetic magnitudes from SDSS spectroscopy (black), Pan-STARRS (blue), and ZTF (purple) data. The observation dates of the bright- and dim-state spectra are marked by vertical blue and orange lines, respectively, in the light curve panels.}
    \label{fig:spectra5}
\end{figure}

\subsection{SDSS-RM sources}\label{sec:RMselection}

For SDSS-RM sources, we use the reprocessed spectroscopic data provided by \citet{Shen2015} and \citet{shen2024sloandigitalskysurvey}. To ensure consistent flux calibration across epochs, we further recalibrate the spectra based on the \Oiii\ $\lambda$5007 flux following the method described in \S\ref{sec:specfitting}.

To identify potential \Heii\ ``changing-look'' events, we visually inspect the root-mean-square (RMS) spectra of all SDSS-RM quasars. Owing to its stronger variability, the \Heii\ $\lambda$4686 broad emission line is generally more prominent in the RMS spectra than in individual single-epoch spectra. Based on this inspection, we select 25 candidate quasars exhibiting the most prominent \Heii\ $\lambda$4686 features in their RMS spectra. We then examine the full set of multi-epoch spectra for each candidate to assess whether the \Heii\ broad emission line shows clear signs of emergence or disappearance. This procedure leads to the identification of three \Heii\ ``CL" quasars: RM17, RM160, and RM772.

For each SDSS-RM source, we perform AGN–host spectral decomposition using the stacked SDSS spectra across all epochs, and subtract the derived host component from each individual-epoch spectrum. The AGN spectral fitting and uncertainty estimation follow the procedures described above.

\begin{figure*}
\hspace{1cm}
    \includegraphics[width=0.9\linewidth]{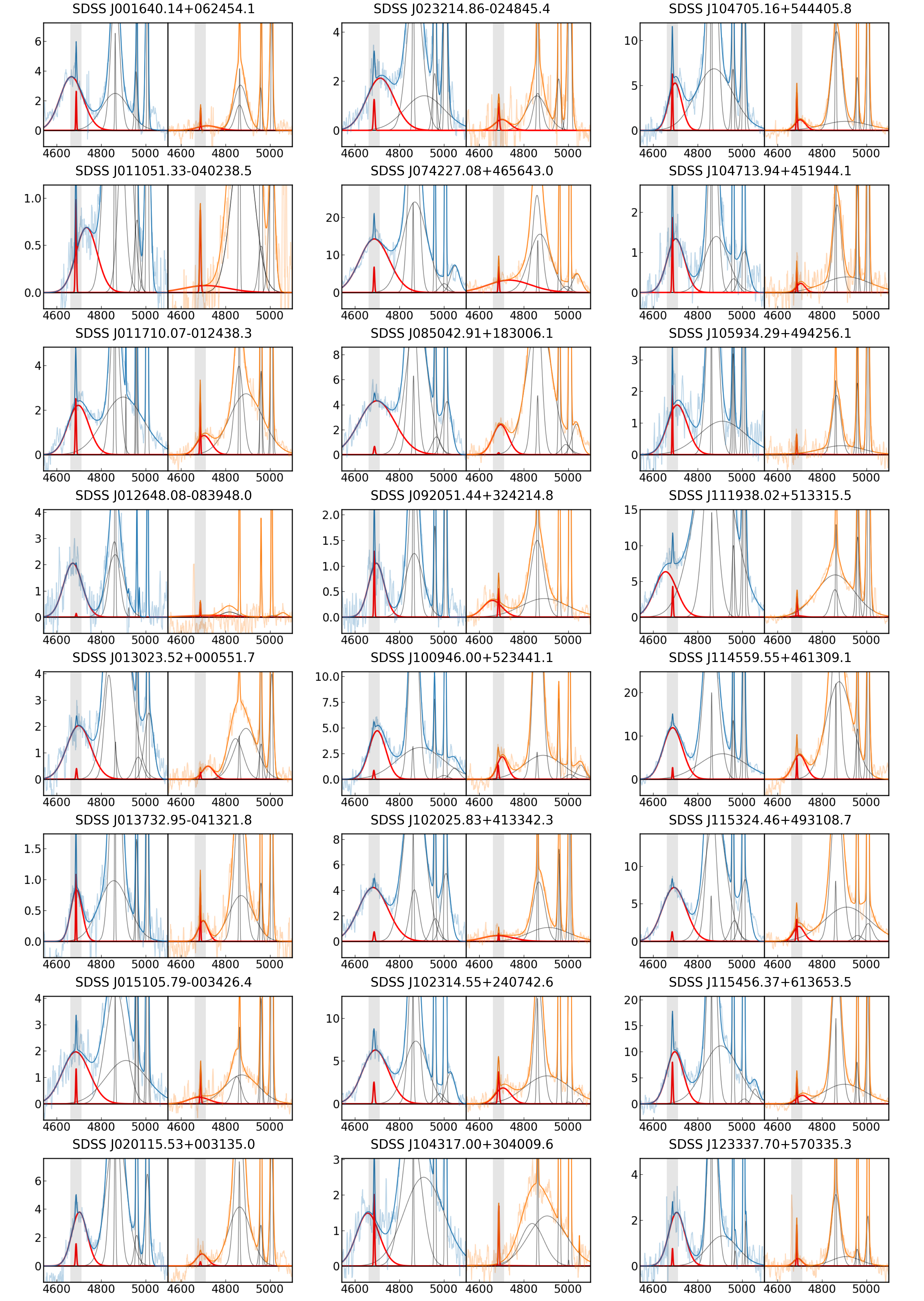}  
\end{figure*}
\begin{figure*}\ContinuedFloat
    \includegraphics[width=0.9\linewidth]{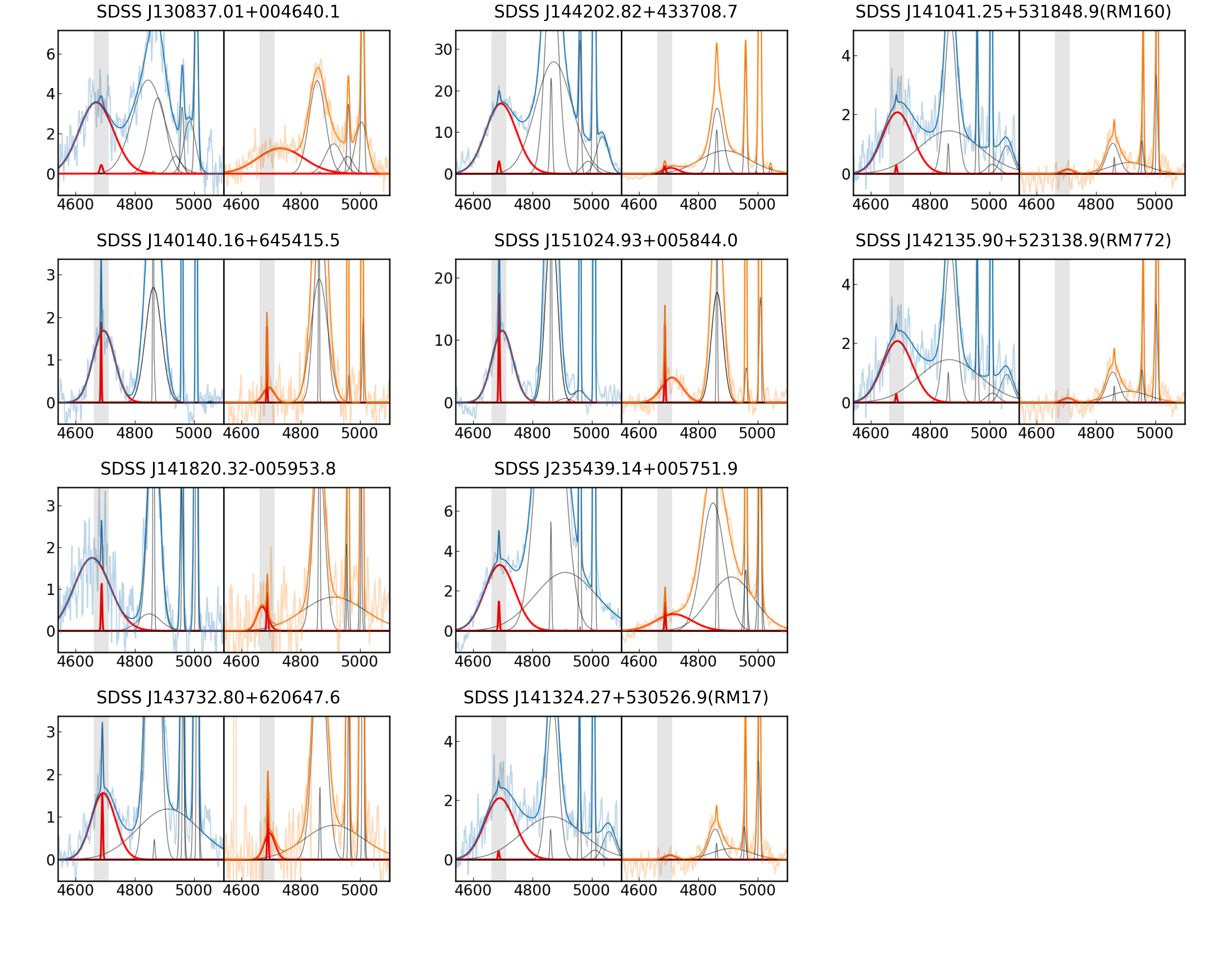}  
    \caption{
Zoom-in view of the \Heii–H$\beta$–\Oiii\ region after subtracting the continuum, host galaxy, and \Feii\ emission. The shaded gray region marks the location of the \Heii\ $\lambda$4686 broad emission line. Each panel shows the spectral fitting results for a \Heii\ ``changing-look" quasar in its bright state (left, blue) and dim state (right, orange). In each subpanel, the observed spectra (faint blue/orange lines) and total best-fit models (thick blue/orange lines) are shown. The best-fit Gaussian components are also shown, with the \Heii\ emission lines highlighted in red, and other features (H$\beta$ and \Oiii) indicated in black.
    \label{fig:lines}}
\end{figure*}

\section{The selected sample} \label{sec:results}

We present in Fig. \ref{fig:spectra1} to \ref{fig:spectra5} the \Heii-bright and \Heii-dim state spectra, along with key auxiliary information, for the 34 \Heii\ ``changing-look" quasars selected in \S\ref{sec:selection} and \S\ref{sec:RMselection} (see appendix for the catalog, and Fig. \ref{fig:lines} for zoomed-in view of the \Heii–H$\beta$–\Oiii\ region). Below each panel in Fig. \ref{fig:spectra1} to \ref{fig:spectra5}, we display the $g$-band light curve for the corresponding source. Most objects exhibit significant continuum variability. We quantify this variability using the maximum $g$-band variation amplitude, $\Delta g_{\mathrm{max}}$, computed from photometric measurements only. We find that 29 of them have $\Delta g_{\mathrm{max}} > 1$ mag, placing them within the population typically classified as extremely variable quasars (EVQs; \citealt{Rumbaugh2018,Ren2022}).

\begin{figure}
\includegraphics[width=1.0\linewidth]{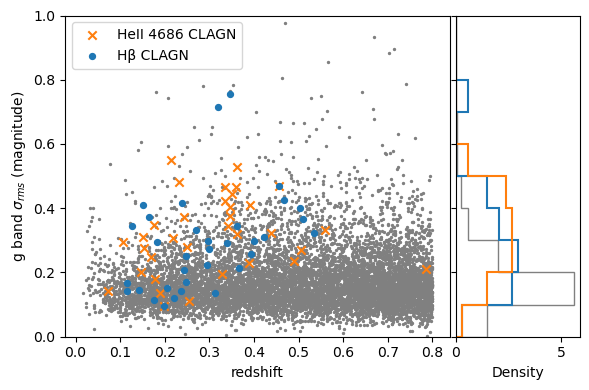}
\caption{The $g$-band variability amplitude ($\sigma_{\mathrm{rms}}$) of our \Heii\ $\lambda$4686 ``CL" quasars (orange) is shown in comparison with that of the parent sample (black, see \S\ref{sec:parentsample}). For reference, the H$\beta$ ``CL" quasars compiled from the literature are also over-plotted (blue). The continuum variability of the H$\beta$ and \Heii\ ``CL" quasars is statistically indistinguishable (p = 0.672, Kolmogorov–Smirnov test), and both populations exhibit significantly stronger variability than the parent sample (p = $9 \times 10^{-10}$ for \Heii\ ``CL" quasars and $2 \times 10^{-6}$ for H$\beta$ ``CL" quasars).
\label{fig:sigmarms}}
\end{figure}

We further calculate the $g$-band variation amplitudes $\sigma_{\mathrm{rms}}$, using the $g$-band photometric light curves, and plot the results in Fig. \ref{fig:sigmarms}. Our \Heii\ ``changing-look" quasars clearly exhibit significantly stronger continuum variability than the parent sample.

We note none of the 34 sources exhibit light curve features characteristic of tidal disruption events (TDEs) \citep[e.g.][]{1988Natur.333..523R,2017ApJ...842...29H,Velzen_2021}. While it is in principle possible for a TDE occurring within an extremely variable quasar (EVQ) to produce the observed \Heii\ variability, such a coincidence would be statistically rare and thus unlikely to account for the majority of our sample. Moreover, \Heii\ broad emission lines (BELs) in TDEs are typically much stronger, often comparable in flux to the H$\beta$ line \citep[e.g.,][]{Charalampopoulos2022}, whereas in our sample the \Heii\ $\lambda$4686 emission remains relatively weak even in the \Heii\ bright state.

\section{discussion} \label{sec:discussion}

\subsection{Comparison with $H\beta$ CLAGN}

\begin{figure}
\includegraphics[width=1.0\linewidth]{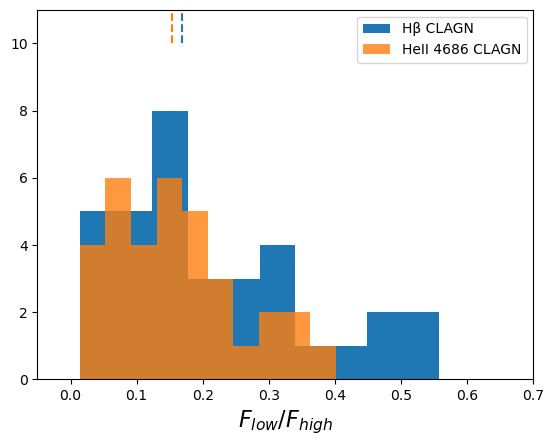}
\caption{Histogram of the broad \Heii\ flux variation amplitude in our \Heii\ “changing-look” quasars, defined as the ratio of line flux in the dim to bright state ($F_{\mathrm{low}}/F_{\mathrm{high}}$). For comparison, we also show the distribution of broad H$\beta$ flux variation amplitudes for H$\beta$ “changing-look” quasars previously identified from SDSS spectroscopy. Vertical lines mark the median values of the two samples. The two distributions are statistically consistent, showing no significant difference in variability amplitude (Kolmogorov–Smirnov test $p = 0.673$).
\label{fig:strength}}
\end{figure}

\begin{figure}
\includegraphics[width=1.0\linewidth]{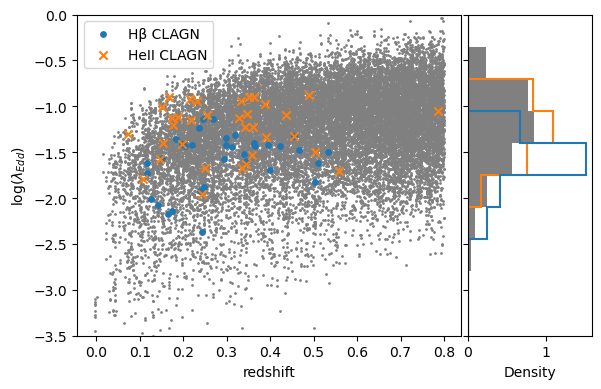}
\caption{The bright-state Eddington ratios as a function of redshift for both \Heii\ and H$\beta$ “changing-look” quasars are shown, with black dots indicating the parent sample defined in \S\ref{sec:parentsample}. The corresponding histograms are presented in the right panel.
\label{fig:Edd}}
\end{figure}

\begin{figure}
\includegraphics[width=1.0\linewidth]{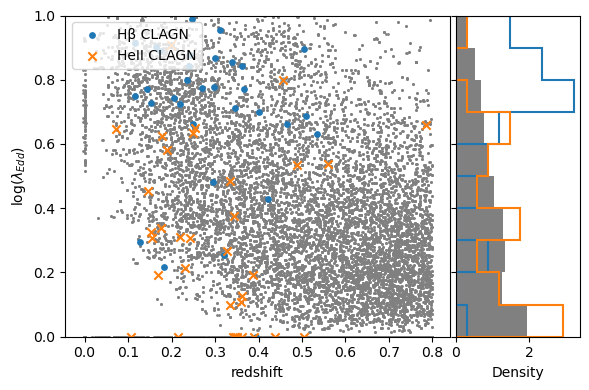}
\caption{The dim-state 5100 \AA\ host fraction as a function of redshift is shown for both \Heii\ and H$\beta$ “changing-look” quasars. Black dots indicate the parent sample defined in \S\ref{sec:parentsample}. The host fraction is defined as the ratio of the host flux to the total flux, both measured within the 5080–5130 \AA\ wavelength range. The corresponding histograms are shown in the right panel.
\label{fig:host}}
\end{figure}

To understand the nature of these \Heii\ ``changing-look" quasars, we compare their properties from multiple perspectives with those of previously identified H$\beta$ ``CL" quasars.

Within our parent sample, we compile all known H$\beta$ “changing-look” quasars identified in the literature based solely on SDSS spectroscopy \citep{Green_2022,2018ApJ...862..109Y,2016MNRAS.457..389M,Potts_2021,LaMassa_2015,Ruan_2016,2018ApJ...858...49W}, yielding a total of 34 objects—a number that coincidentally matches the count of \Heii\ ``CL" quasars in this study. For these sources, we determine the broad H$\beta$ bright and dim states as the epochs with the highest and lowest broad H$\beta$ fluxes, respectively, based on our own spectral fitting. This ensures a consistent comparison with the \Heii\ “changing-look” quasar sample constructed in this work.

We find that the flux variation amplitudes of the broad \Heii\ lines in \Heii\ ``CL" quasars are statistically comparable to those of the broad H$\beta$ lines in H$\beta$ ``CL" quasars (Fig.\ref{fig:strength}). In fact, the \Heii\ ``CL" quasars exhibit a slightly stronger median variation amplitude, as indicated by the vertical dashed lines in Fig.\ref{fig:strength}. This result suggests that, in terms of broad emission-line variability, our \Heii\ ``CL" quasars resemble the known H$\beta$ ``CL" quasars. Therefore, it is reasonable to classify the sources we selected as \Heii\ ``changing-look" quasars in an analogous way.

We further compare the $g$-band variability amplitudes of the two populations (Fig.~\ref{fig:sigmarms}). Most sources in both samples exhibit significant $g$-band variability, implying that the observed ``changing-look" behavior in both broad \Heii\ and H$\beta$ is primarily driven by substantial variations in the ionizing continuum.

In Fig.~\ref{fig:Edd}, we show the Eddington ratios of \Heii\ ``CL" quasars as a function of redshift, in comparison with those the parent sample (see \S\ref{sec:parentsample}), and of the H$\beta$ ``CL" sample. To estimate the black hole mass, we adopt the single-epoch (based on the corresponding bright-state spectra) virial formula from \cite{shen2024sloandigitalskysurvey}: $Log{M} =0.5Log(L_{5100}) +2Log(FWHM_{H\beta}) +0.85$, and compute the bolometric luminosity as $9.26L_{5100}$ \citep{2011ApJS..194...45S}. We note that the systematic uncertainty in single-epoch mass estimates is typically around 0.5 dex in $\log M$, and we therefore focus only on the overall distribution rather than detailed comparisons of individual sources.

As previously reported in the literature \citep[e.g.][]{Green_2022,MacLeod_2019,wang2024identifyingchanginglookagnsusing}, H$\beta$ ``CL" quasars tend to have lower Eddington ratios compared to the parent sample. This trend is also evident in Fig.~\ref{fig:Edd} (KS test p-value =$4 \times 10^{-8}$ ). In contrast, \Heii\ ``CL" quasars show relatively higher Eddington ratios than H$\beta$ ``CL" quasars (KS test p-value =0.0001), while their distribution is statistically consistent with that of the parent sample (KS test p-value = 0.097).

This suggests two key points. First, \Heii\ ``CL" transitions can occur at higher Eddington ratios than H$\beta$ ``CL" events. A likely explanation is that the ionizing continuum responsible for \Heii\ originates from more energetic (shorter-wavelength) extreme-UV photons, which are known to show larger variability amplitudes \citep{Welsh2011,Zhu2016}. Second, our selection of \Heii\ ``CL" quasars requires the broad \Heii\ line to be visually detectable in the bright-state spectrum. Strong host-galaxy contamination makes this difficult for low-Eddington-ratio quasars, introducing a selection bias toward higher-Eddington-ratio sources.

We also compared the dim-state host contamination among the \Heii\ ``CL" quasars in Fig.~\ref{fig:host}, the parent sample, and the H$\beta$ ``CL" quasars. We find that H$\beta$ ``CL" quasars exhibit significantly higher host contamination fractions than the parent sample (KS test p-value = $5 \times 10^{-13}$), consistent with their systematically lower Eddington ratios. In contrast, the host contamination in \Heii\ ``CL" quasars is substantially lower than that in H$\beta$ ``CL" quasars (p-value = $3 \times 10^{-9}$), and statistically comparable to that of the parent sample (p-value = 0.553).

\subsection{Comparison of the Variability of Broad \Heii\ and Broad H$\beta$}

\begin{figure}
\includegraphics[width=1.0\linewidth]{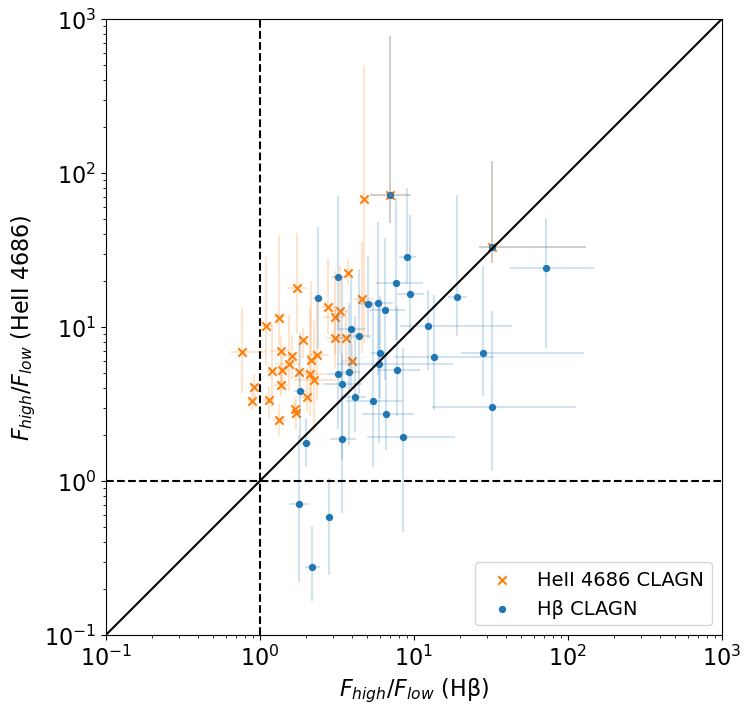}
\caption{The broad \Heii\ flux variability amplitude (between the bright and dim states) is plotted against that of the broad H$\beta$ for our \Heii\ ``CL" quasars, with H$\beta$ ``CL" quasars over-plotted for comparison.
To mitigate large statistical fluctuations caused by the typically much weaker \Heii\ line in the dim state, both variability amplitudes are derived from the median values of Monte Carlo simulations. The corresponding uncertainties represent the 68\% confidence intervals.
\label{fig:ratio}}
\end{figure}

In Fig.~\ref{fig:ratio}, we compare the flux variability amplitudes of broad \Heii\ and broad H$\beta$ (between the bright and dim states) for our \Heii\ ``CL" quasars, with H$\beta$ ``CL" quasars over-plotted for comparison. We find that \Heii\ ``CL" quasars generally exhibit more prominent \Heii\ variability than H$\beta$. While this is partly expected given our selection based on strong \Heii\ flux changes, it is noteworthy that the \Heii\ and H$\beta$ variability amplitudes are comparable in H$\beta$ ``CL" quasars. This suggests that, in general, \Heii\ is intrinsically more variable than H$\beta$ in quasars.

This trend is consistent with the fact that \Heii\ lines exhibit stronger variability than other broad lines in the rms spectra of SDSS-RM quasars \citep{shen2024sloandigitalskysurvey}. Such behavior can be naturally explained by the stronger variability of higher-energy extreme-UV ionizing photons in quasars \citep{Welsh2011,Zhu2016}, to which \Heii\ is more sensitive due to its much higher ionization potential (four times that of Hydrogen). Moreover, the broad \Heii\ emission is expected to arise from a more compact region closer to the ionizing source \citep[e.g.][]{2012ApJ...744L...4G,2017ApJ...846...79L,2010ApJ...716..993B}, which leads to a weaker smoothing effect in its response to continuum variations.

Interestingly, the above comparison between \Heii\ and H$\beta$ is in some sense analogous to that between H$\beta$ and \Mgii: it is well known that \Mgii\ is less variable than the Balmer broad emission lines \citep[e.g.,][]{shen2024sloandigitalskysurvey,2015ApJ...811...42S}. This trend is also reflected in H$\beta$ ``CL" quasars, where \Mgii\ often exhibits only mild changes and frequently remains detectable even in the dim state \citep[e.g.,][]{Guo_2019,2016MNRAS.457..389M,10.1093/mnras/staa645}.

To explain the connection between H$\beta$ ``CL" and \Mgii\ CL, 
\citet{Guo2020}
proposed a ``CL" sequence in which, as the continuum luminosity gradually declines, H$\beta$ disappears first, followed by H$\alpha$, and eventually \Mgii. In this framework, \Heii\ $\lambda4686$ may also be incorporated into the sequence, preceding H$\beta$ in its disappearance due to its significantly higher ionization potential.

\subsection{The occurrence rate  and selection bias}

We note that there is minimal overlap between our \Heii\ “CL” quasars and the previously identified H$\beta$ “CL” quasars—only two sources are shared: SDSS J012648.08$-$083948.0 and one RM quasar, SDSS J141324.27+530526.9 (RM17). Below, we examine the reasons for this limited overlap.

Among the 32 known H$\beta$ “CL” quasars that were not selected as \Heii\ “CL” quasars, 21 fail our \Heii\ selection due to weak broad \Heii\ $\lambda$4686 emission (i.e., equivalent width $<$ 10 \AA), 6 others do not meet the threshold for \Heii\ flux variability (flux change $<$ 2.5 times or S/N $<$ 3), and five lack a visually distinct \Heii\ bump near 4686 \AA, likely due to strong \Feii\ contamination or insufficient S/N, and were excluded during visual inspection.

These findings imply that a significant fraction of H$\beta$ “CL” quasars may exhibit \Heii\ variability as well, though weak \Heii\ emission or insufficient data quality currently limits their detectability with our selection methods.

We estimate the occurrence rate of \Heii\ “CL” quasars as follows. Excluding the 3 RM quasars, the remaining 31 \Heii\ “CL” quasars are identified from a parent sample of 9,363 quasars, resulting in a raw occurrence rate of approximately 0.3\%. However, many quasars in the parent sample exhibit very weak broad \Heii\ emission. Only 6,227 quasars have a \Heii\ EW greater than 10 \AA, satisfying our additional selection criterion. Taking this into account, we estimate the intrinsic occurrence rate to be about 0.5\%. Given the heavy blending of \Heii\ $\lambda$4686 with \Feii\ and the exclusion of numerous candidates during visual inspection, the actual occurrence rate of \Heii\ “CL” quasars is likely underestimated and could be substantially higher.

\begin{figure}
\includegraphics[width=1.0\linewidth]{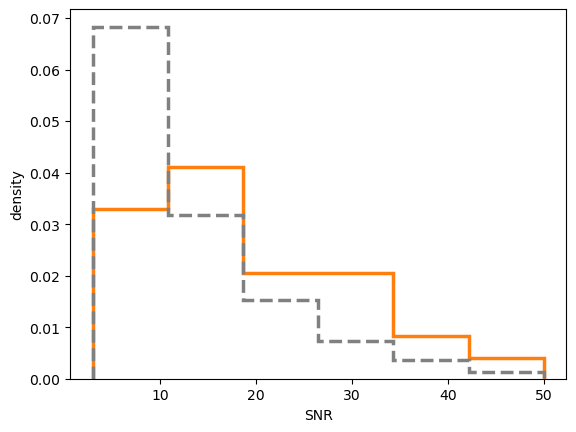}
\caption{The distribution of high-state spectral S/N ratios is shown for \Heii\ “CL” quasars (yellow solid line, compared to the parent sample (grey dotted line) constructed in \S\ref{sec:parentsample}.
\label{fig:SNhistogram}}
\end{figure}

The identification of \Heii\ “CL” quasars is also sensitive to spectral S/N. In Fig.~\ref{fig:SNhistogram}, we compare the S/N distributions of the high-state spectra for the \Heii\ “changing-look” quasars and the parent sample. The occurrence rate clearly increases with S/N. Restricting to spectra with S/N $>15$, we find 18 \Heii\ “CL” quasars among 2,157 parent sample quasars, corresponding to an occurrence rate of approximately 0.8\%. 

The apparent occurrence rate of H$\beta$ “CL” (CL) quasars in our parent sample (33 out of 9,363, excluding RM sources) is quite similar to that of \Heii\ “CL” quasars (31 out of 9,363). Reported occurrence rates of H$\beta$ CL events in recent literature are also comparable. For instance,  
\citet{Guo_2025} reported a rate of 0.56\%, while \citet{wang2024identifyingchanginglookagnsusing} estimated a range of 0.3\% to 0.7\%.

We also note that the occurrence rates of both \Heii\ and H$\beta$ ``CL" events depend on redshift. At $z = 0.6$–$0.8$, such events are nearly absent, which may be due to the higher luminosities of quasars at higher redshifts, resulting in intrinsically weaker variability \citep[e.g.,][]{Zuo_2012,Zeltyn_2024,Chanchaiworawit_2024}.

We conclude by discussing the selection biases involved in identifying ``CL" events. For \Heii, which is intrinsically weak and heavily blended with \Feii, the identification of ``CL" events is primarily limited by the requirement of visually detecting a broad \Heii\ line during the bright state. As a result, the selection is biased against quasars with intrinsically weak \Heii\ emission, low spectral signal-to-noise ratio (S/N), and strong host-galaxy contamination. We expect the number of detectable \Heii\ ``CL" events to significantly increase with higher spectral S/N.

In contrast, the selection bias for H$\beta$ (and other prominent lines) ``CL" quasars operates in the opposite direction. Since H$\beta$ is typically strong, previous studies have often struggled to find dim-state spectra in which the broad H$\beta$ line is genuinely absent, resulting in considerable but arguably unnecessary effort devoted to visually confirming such cases\footnote{A similar challenge is present in this work, where considerable effort is required to visually confirm the presence of broad \Heii\ emission in the bright state, given its weakness and blending with nearby features.}. However, in practice, broad H$\beta$ remains statistically detectable in the dim-state spectra of most reported H$\beta$ ``CL" quasars \citep[e.g.][]{Ren_2024}. Therefore, improving the S/N of the spectra could actually reduce the number of H$\beta$ ``CL" identifications by revealing residual broad-line features in the dim state.

These selection effects fundamentally arise from the commonly adopted definition of changing-look events, which typically requires the emergence or disappearance of a broad emission line. While this definition has proven useful in identifying dramatic spectral changes, it is not mathematically rigorous and may introduce complexities in sample selection. From this perspective, it may be more informative and straightforward to study quasars exhibiting significant variations in broad emission lines and the continuum—particularly the ionizing continuum—even if they do not strictly meet the traditional ``CL" definition. Such an approach could help mitigate selection biases and provide a more comprehensive view of the underlying physical processes.

\section{Summary}\label{sec:summary}

In this work, we conduct the first systematic investigation of changing-look (CL) behavior in the broad \Heii\ $\lambda$4686 emission line using repeated spectroscopy from SDSS. We identify 34 \Heii\ “CL” quasars that show dramatic changes in broad \Heii\ emission between spectroscopic epochs.

Our key findings are as follows:

1. The broad \Heii\ line variability in these \Heii\ ``CL" quasars, measured between the bright and dim states, is comparably large in amplitude to that of broad H$\beta$ in known H$\beta$ “CL” quasars  compiled in the same parent sample.

2. Compared to the parent sample, both \Heii\ and H$\beta$ “CL” quasars display significantly enhanced g-band photometric variability amplitudes, indicating that large-scale changes in the continuum are a key driver of the ``CL" events.

3.  \Heii\ ``CL" quasars tend to show systematically higher Eddington ratios and lower host-galaxy contamination fractions than their H$\beta$ counterparts, very likely driven by selection effects.

4. In \Heii\ ``CL" quasars, the broad \Heii\ line flux varies more than H$\beta$, while in H$\beta$ ``CL" quasars, \Heii\ variability is comparable to that of H$\beta$. This suggests that \Heii\ intrinsically varies more, likely driven by the more variable extreme ultraviolet (EUV) continuum.

5. The estimated occurrence rate of \Heii\ CL quasars is approximately 0.5\%, but this rate is sensitive to various observational and physical factors, including spectral signal-to-noise ratio (S/N), \Heii\ line equivalent width (EW), blending with \Feii, Eddington ratio, host galaxy contamination, and redshift.

6. There is minimal overlap between \Heii\ and H$\beta$ ``CL" quasars identified in the same parent sample, primarily due to selection biases introduced by the traditional definition of ``CL"—namely, the ``appearance" or ``disappearance" of a specific broad emission line.

Collectively, these findings demonstrate that broad \Heii\ emission is a valuable probe of extreme variability in AGNs and emphasize the need to account for selection biases in “changing-look” studies.

\begin{acknowledgments}
The work of JXW is supported by National Natural Science Foundation of China (grant nos. 12033006, 12192221) and the Cyrus Chun Ying Tang Foundations.

\end{acknowledgments}

\facilities{SDSS, Panstarr, ZTF}

\software{Astropy \citep{astropy:2013, astropy:2018, astropy:2022},
          PyQSOFit \citep{2018ascl.soft09008G,2019ApJS..241...34S,2024ApJ...974..153R}
          }

\appendix

\begin{deluxetable}{lccccc}
    \tablecaption{The sample of 34 \Heii\ 4686 ``CL" quasars identified in this work. 
    The observation dates (MJD) of the spectra in the bright and dim states of the broad \Heii\ line, along with the corresponding broad \Heii\ line fluxes (in unit of $10^{-17} erg/s/cm^2$), are presented.  
    }
    \label{tab:my_label}
    \tablehead{
    \colhead{Source Name} & \colhead{Redshift} & \colhead{MJD$_{bright}$} & \colhead{\Heii\ flux$_{bright}$} & \colhead{MJD$_{dim}$} & \colhead{\Heii\ flux$_{dim}$}
    }
    \startdata
SDSS J001640.14+062454.1&0.177&58429&$298.85^{+111.41}_{-44.01}$&58124&$27.17^{+18.86}_{-12.85}$\\
SDSS J011051.33-040238.5&0.504&57724&$72.30^{+16.58}_{-18.94}$&57309&$7.30^{+4.02}_{-4.78}$\\
SDSS J011710.07-012438.3&0.388&58072&$285.08^{+38.86}_{-29.43}$&56980&$68.75^{+9.26}_{-9.48}$\\
SDSS J012648.08-083948.0&0.198&52163&$222.67^{+34.96}_{-29.72}$&54465&$7.21^{+0.09}_{-5.35}$\\
SDSS J013023.52+000551.7&0.345&51817&$246.31^{+57.81}_{-41.14}$&56989&$29.09^{+13.92}_{-9.50}$\\
SDSS J013732.95-041321.8&0.437&57339&$56.03^{+7.89}_{-7.54}$&57711&$20.19^{+4.57}_{-4.13}$\\
SDSS J015105.79-003426.4&0.335&51793&$328.68^{+47.49}_{-52.76}$&56982&$25.23^{+22.59}_{-12.10}$\\
SDSS J020115.53+003135.0&0.362&51812&$357.46^{+41.08}_{-38.20}$&58079&$54.70^{+20.96}_{-13.63}$\\
SDSS J023214.86-024845.4&0.785&57727&$272.76^{+87.68}_{-133.51}$&55557&$56.75^{+41.75}_{-31.78}$\\
SDSS J074227.08+465643.0&0.168&58429&$2440.20^{+770.88}_{-150.86}$&53317&$741.83^{+122.64}_{-270.18}$\\
SDSS J085042.91+183006.1&0.328&53708&$852.64^{+86.08}_{-93.47}$&58138&$207.65^{+29.09}_{-30.27}$\\
SDSS J092051.44+324214.8&0.342&56325&$88.80^{+10.98}_{-12.38}$&54550&$35.14^{+8.16}_{-6.16}$\\
SDSS J100946.00+523441.1&0.175&52385&$456.83^{+45.26}_{-34.73}$&52400&$136.58^{+39.67}_{-20.75}$\\
SDSS J102025.83+413342.3&0.361&53002&$629.02^{+71.65}_{-104.65}$&58140&$71.32^{+23.13}_{-22.81}$\\
SDSS J102314.55+240742.6&0.188&56309&$886.56^{+115.98}_{-166.03}$&57866&$167.30^{+31.35}_{-28.71}$\\
SDSS J104317.00+304009.6&0.243&58132&$169.79^{+27.07}_{-30.53}$&56365&$28.38^{+20.24}_{-20.11}$\\
SDSS J104705.16+544405.8&0.215&52368&$393.85^{+53.88}_{-41.28}$&57429&$65.97^{+11.95}_{-11.28}$\\
SDSS J104713.94+451944.1&0.489&53053&$130.63^{+29.81}_{-20.00}$&56992&$20.29^{+17.81}_{-10.79}$\\
SDSS J105934.29+494256.1&0.346&52669&$163.64^{+30.63}_{-33.71}$&57127&$10.41^{+12.81}_{-5.86}$\\
SDSS J111938.02+513315.5&0.107&52353&$793.86^{+91.00}_{-95.80}$&57071&$11.99^{+87.67}_{-10.45}$\\
SDSS J114559.55+461309.1&0.154&56776&$1306.15^{+51.32}_{-47.82}$&53054&$397.17^{+53.38}_{-67.43}$\\
SDSS J115324.46+493108.7&0.334&57134&$930.13^{+56.25}_{-149.83}$&56385&$124.08^{+87.04}_{-17.62}$\\
SDSS J115456.37+613653.5&0.152&52320&$919.62^{+92.01}_{-91.13}$&56663&$111.05^{+25.31}_{-15.61}$\\
SDSS J123337.70+570335.3&0.351&52790&$218.08^{+27.82}_{-26.95}$&56426&$16.18^{+7.55}_{-8.49}$\\
SDSS J130837.01+004640.1&0.559&51585&$579.45^{+136.81}_{-94.13}$&51985&$117.30^{+97.31}_{-73.15}$\\
SDSS J140140.16+645415.5&0.146&56717&$151.12^{+15.49}_{-16.68}$&51984&$29.29^{+18.46}_{-14.72}$\\
SDSS J141820.32-005953.8&0.254&51957&$280.97^{+44.21}_{-45.64}$&51609&$40.35^{+30.96}_{-18.77}$\\
SDSS J143732.80+620647.6&0.219&56419&$190.31^{+26.52}_{-33.90}$&52368&$32.48^{+14.03}_{-15.96}$\\
SDSS J144202.82+433708.7&0.231&52734&$2207.77^{+122.99}_{-132.04}$&58161&$98.24^{+15.56}_{-17.82}$\\
SDSS J151024.93+005844.0&0.072&51665&$1017.90^{+43.28}_{-44.83}$&51996&$346.52^{+46.04}_{-44.41}$\\
SDSS J235439.14+005751.9&0.390&57682&$553.84^{+90.94}_{-125.86}$&56959&$110.13^{+14.80}_{-45.74}$\\
SDSS J141324.27+530526.9(RM17)&0.456&57550&$417.21^{+72.21}_{-85.25}$&56664&$5.93^{+1.35}_{-5.39}$\\
SDSS J141041.25+531848.9(RM160)&0.359&57576&$71.21^{+6.77}_{-8.18}$&58587&$6.16^{+2.87}_{-4.29}$\\
SDSS J142135.90+523138.9(RM772)&0.249&58216&$257.82^{+49.39}_{-40.93}$&56664&$14.69^{+13.55}_{-7.59}$\\
\enddata

\end{deluxetable}

\bibliography{sample7__1_}{}

\begin{thebibliography}{}
\expandafter\ifx\csname natexlab\endcsname\relax\def\natexlab#1{#1}\fi
\providecommand{\url}[1]{\href{#1}{#1}}
\providecommand{\dodoi}[1]{doi:~\href{http://doi.org/#1}{\nolinkurl{#1}}}
\providecommand{\doeprint}[1]{\href{http://ascl.net/#1}{\nolinkurl{http://ascl.net/#1}}}
\providecommand{\doarXiv}[1]{\href{https://arxiv.org/abs/#1}{\nolinkurl{https://arxiv.org/abs/#1}}}

\bibitem[{Abdurro’uf {et~al.}(2022)Abdurro’uf, Accetta, Aerts, Silva~Aguirre, Ahumada, Ajgaonkar, Filiz~Ak, Alam, Allende~Prieto, Almeida, Anders, Anderson, Andrews, Anguiano, Aquino-Ortíz, Aragón-Salamanca, Argudo-Fernández, Ata, Aubert, Avila-Reese, Badenes, Barbá, Barger, Barrera-Ballesteros, Beaton, Beers, Belfiore, Bender, Bernardi, Bershady, Beutler, Bidin, Bird, Bizyaev, Blanc, Blanton, Boardman, Bolton, Boquien, Borissova, Bovy, Brandt, Brown, Brownstein, Brusa, Buchner, Bundy, Burchett, Bureau, Burgasser, Cabang, Campbell, Cappellari, Carlberg, Wanderley, Carrera, Cash, Chen, Chen, Cherinka, Chiappini, Choi, Chojnowski, Chung, Clerc, Cohen, Comerford, Comparat, da~Costa, Covey, Crane, Cruz-Gonzalez, Culhane, Cunha, Dai, Damke, Darling, Davidson~Jr., Davies, Dawson, De~Lee, Diamond-Stanic, Cano-Díaz, Sánchez, Donor, Duckworth, Dwelly, Eisenstein, Elsworth, Emsellem, Eracleous, Escoffier, Fan, Farr, Feng, Fernández-Trincado, Feuillet, Filipp, Fillingham, Frinchaboy, Fromenteau, Galbany,
  García, García-Hernández, Ge, Geisler, Gelfand, Géron, Gibson, Goddy, Godoy-Rivera, Grabowski, Green, Greener, Grier, Griffith, Guo, Guy, Hadjara, Harding, Hasselquist, Hayes, Hearty, Hernández, Hill, Hogg, Holtzman, Horta, Hsieh, Hsu, Hsu, Huber, Huertas-Company, Hutchinson, Hwang, Ibarra-Medel, Chitham, Ilha, Imig, Jaekle, Jayasinghe, Ji, Johnson, Jones, Jönsson, Katkov, Khalatyan, Kinemuchi, Kisku, Knapen, Kneib, Kollmeier, Kong, Kounkel, Kreckel, Krishnarao, Lacerna, Lane, Langgin, Lavender, Law, Lazarz, Leung, Leung, Lewis, Li, Li, Lian, Liang, Lin, Lin, Lin, Lintott, Long, Longa-Peña, López-Cobá, Lu, Lundgren, Luo, Mackereth, de~la Macorra, Mahadevan, Majewski, Manchado, Mandeville, Maraston, Margalef-Bentabol, Masseron, Masters, Mathur, McDermid, Mckay, Merloni, Merrifield, Meszaros, Miglio, Di~Mille, Minniti, Minsley, Monachesi, Moon, Mosser, Mulchaey, Muna, Muñoz, Myers, Myers, Nadathur, Nair, Nandra, Neumann, Newman, Nidever, Nikakhtar, Nitschelm, O’Connell, Garma-Oehmichen, Luan
  Souza~de Oliveira, Olney, Oravetz, Ortigoza-Urdaneta, Osorio, Otter, Pace, Padilla, Pan, Pan, Parikh, Parker, Peirani, Peña~Ramírez, Penny, Percival, Perez-Fournon, Pinsonneault, Poidevin, Poovelil, Price-Whelan, Bárbara~de Andrade~Queiroz, Raddick, Ray, Rembold, Riddle, Riffel, Riffel, Rix, Robin, Rodríguez-Puebla, Roman-Lopes, Román-Zúñiga, Rose, Ross, Rossi, Rubin, Salvato, Sánchez, Sánchez-Gallego, Sanderson, Santana~Rojas, Sarceno, Sarmiento, Sayres, Sazonova, Schaefer, Schiavon, Schlegel, Schneider, Schultheis, Schwope, Serenelli, Serna, Shao, Shapiro, Sharma, Shen, Shetrone, Shu, Simon, Skrutskie, Smethurst, Smith, Sobeck, Spoo, Sprague, Stark, Stassun, Steinmetz, Stello, Stone-Martinez, Storchi-Bergmann, Stringfellow, Stutz, Su, Taghizadeh-Popp, Talbot, Tayar, Telles, Teske, Thakar, Theissen, Tkachenko, Thomas, Tojeiro, Hernandez~Toledo, Troup, Trump, Trussler, Turner, Tuttle, Unda-Sanzana, Vázquez-Mata, Valentini, Valenzuela, Vargas-González, Vargas-Magaña, Alfaro, Villanova, Vincenzo,
  Wake, Warfield, Washington, Weaver, Weijmans, Weinberg, Weiss, Westfall, Wild, Wilde, Wilson, Wilson, Wilson, Wolf, Wood-Vasey, Yan, Zamora, Zasowski, Zhang, Zhao, Zheng, Zheng, \& Zhu}]{Abdurrouf_2022}
Abdurro’uf, Accetta, K., Aerts, C., {et~al.} 2022, The Astrophysical Journal Supplement Series, 259, 35, \dodoi{10.3847/1538-4365/ac4414}

\bibitem[{{Antonucci}(1993)}]{1993ARA&A..31..473A}
{Antonucci}, R. 1993, \araa, 31, 473, \dodoi{10.1146/annurev.aa.31.090193.002353}

\bibitem[{{Astropy Collaboration} {et~al.}(2013){Astropy Collaboration}, {Robitaille}, {Tollerud}, {Greenfield}, {Droettboom}, {Bray}, {Aldcroft}, {Davis}, {Ginsburg}, {Price-Whelan}, {Kerzendorf}, {Conley}, {Crighton}, {Barbary}, {Muna}, {Ferguson}, {Grollier}, {Parikh}, {Nair}, {Unther}, {Deil}, {Woillez}, {Conseil}, {Kramer}, {Turner}, {Singer}, {Fox}, {Weaver}, {Zabalza}, {Edwards}, {Azalee Bostroem}, {Burke}, {Casey}, {Crawford}, {Dencheva}, {Ely}, {Jenness}, {Labrie}, {Lim}, {Pierfederici}, {Pontzen}, {Ptak}, {Refsdal}, {Servillat}, \& {Streicher}}]{astropy:2013}
{Astropy Collaboration}, {Robitaille}, T.~P., {Tollerud}, E.~J., {et~al.} 2013, \aap, 558, A33, \dodoi{10.1051/0004-6361/201322068}

\bibitem[{{Astropy Collaboration} {et~al.}(2018){Astropy Collaboration}, {Price-Whelan}, {Sip{\H{o}}cz}, {G{\"u}nther}, {Lim}, {Crawford}, {Conseil}, {Shupe}, {Craig}, {Dencheva}, {Ginsburg}, {Vand erPlas}, {Bradley}, {P{\'e}rez-Su{\'a}rez}, {de Val-Borro}, {Aldcroft}, {Cruz}, {Robitaille}, {Tollerud}, {Ardelean}, {Babej}, {Bach}, {Bachetti}, {Bakanov}, {Bamford}, {Barentsen}, {Barmby}, {Baumbach}, {Berry}, {Biscani}, {Boquien}, {Bostroem}, {Bouma}, {Brammer}, {Bray}, {Breytenbach}, {Buddelmeijer}, {Burke}, {Calderone}, {Cano Rodr{\'\i}guez}, {Cara}, {Cardoso}, {Cheedella}, {Copin}, {Corrales}, {Crichton}, {D'Avella}, {Deil}, {Depagne}, {Dietrich}, {Donath}, {Droettboom}, {Earl}, {Erben}, {Fabbro}, {Ferreira}, {Finethy}, {Fox}, {Garrison}, {Gibbons}, {Goldstein}, {Gommers}, {Greco}, {Greenfield}, {Groener}, {Grollier}, {Hagen}, {Hirst}, {Homeier}, {Horton}, {Hosseinzadeh}, {Hu}, {Hunkeler}, {Ivezi{\'c}}, {Jain}, {Jenness}, {Kanarek}, {Kendrew}, {Kern}, {Kerzendorf}, {Khvalko}, {King}, {Kirkby}, {Kulkarni},
  {Kumar}, {Lee}, {Lenz}, {Littlefair}, {Ma}, {Macleod}, {Mastropietro}, {McCully}, {Montagnac}, {Morris}, {Mueller}, {Mumford}, {Muna}, {Murphy}, {Nelson}, {Nguyen}, {Ninan}, {N{\"o}the}, {Ogaz}, {Oh}, {Parejko}, {Parley}, {Pascual}, {Patil}, {Patil}, {Plunkett}, {Prochaska}, {Rastogi}, {Reddy Janga}, {Sabater}, {Sakurikar}, {Seifert}, {Sherbert}, {Sherwood-Taylor}, {Shih}, {Sick}, {Silbiger}, {Singanamalla}, {Singer}, {Sladen}, {Sooley}, {Sornarajah}, {Streicher}, {Teuben}, {Thomas}, {Tremblay}, {Turner}, {Terr{\'o}n}, {van Kerkwijk}, {de la Vega}, {Watkins}, {Weaver}, {Whitmore}, {Woillez}, {Zabalza}, \& {Astropy Contributors}}]{astropy:2018}
{Astropy Collaboration}, {Price-Whelan}, A.~M., {Sip{\H{o}}cz}, B.~M., {et~al.} 2018, \aj, 156, 123, \dodoi{10.3847/1538-3881/aabc4f}

\bibitem[{{Astropy Collaboration} {et~al.}(2022){Astropy Collaboration}, {Price-Whelan}, {Lim}, {Earl}, {Starkman}, {Bradley}, {Shupe}, {Patil}, {Corrales}, {Brasseur}, {N{"o}the}, {Donath}, {Tollerud}, {Morris}, {Ginsburg}, {Vaher}, {Weaver}, {Tocknell}, {Jamieson}, {van Kerkwijk}, {Robitaille}, {Merry}, {Bachetti}, {G{"u}nther}, {Aldcroft}, {Alvarado-Montes}, {Archibald}, {B{'o}di}, {Bapat}, {Barentsen}, {Baz{'a}n}, {Biswas}, {Boquien}, {Burke}, {Cara}, {Cara}, {Conroy}, {Conseil}, {Craig}, {Cross}, {Cruz}, {D'Eugenio}, {Dencheva}, {Devillepoix}, {Dietrich}, {Eigenbrot}, {Erben}, {Ferreira}, {Foreman-Mackey}, {Fox}, {Freij}, {Garg}, {Geda}, {Glattly}, {Gondhalekar}, {Gordon}, {Grant}, {Greenfield}, {Groener}, {Guest}, {Gurovich}, {Handberg}, {Hart}, {Hatfield-Dodds}, {Homeier}, {Hosseinzadeh}, {Jenness}, {Jones}, {Joseph}, {Kalmbach}, {Karamehmetoglu}, {Ka{l}uszy{'n}ski}, {Kelley}, {Kern}, {Kerzendorf}, {Koch}, {Kulumani}, {Lee}, {Ly}, {Ma}, {MacBride}, {Maljaars}, {Muna}, {Murphy}, {Norman}, {O'Steen},
  {Oman}, {Pacifici}, {Pascual}, {Pascual-Granado}, {Patil}, {Perren}, {Pickering}, {Rastogi}, {Roulston}, {Ryan}, {Rykoff}, {Sabater}, {Sakurikar}, {Salgado}, {Sanghi}, {Saunders}, {Savchenko}, {Schwardt}, {Seifert-Eckert}, {Shih}, {Jain}, {Shukla}, {Sick}, {Simpson}, {Singanamalla}, {Singer}, {Singhal}, {Sinha}, {Sip{H{o}}cz}, {Spitler}, {Stansby}, {Streicher}, {{{S}}umak}, {Swinbank}, {Taranu}, {Tewary}, {Tremblay}, {Val-Borro}, {Van Kooten}, {Vasovi{'c}}, {Verma}, {de Miranda Cardoso}, {Williams}, {Wilson}, {Winkel}, {Wood-Vasey}, {Xue}, {Yoachim}, {Zhang}, {Zonca}, \& {Astropy Project Contributors}}]{astropy:2022}
{Astropy Collaboration}, {Price-Whelan}, A.~M., {Lim}, P.~L., {et~al.} 2022, \apj, 935, 167, \dodoi{10.3847/1538-4357/ac7c74}

\bibitem[{Bellm {et~al.}(2018)Bellm, Kulkarni, Graham, Dekany, Smith, Riddle, Masci, Helou, Prince, Adams, Barbarino, Barlow, Bauer, Beck, Belicki, Biswas, Blagorodnova, Bodewits, Bolin, Brinnel, Brooke, Bue, Bulla, Burruss, Cenko, Chang, Connolly, Coughlin, Cromer, Cunningham, De, Delacroix, Desai, Duev, Eadie, Farnham, Feeney, Feindt, Flynn, Franckowiak, Frederick, Fremling, Gal-Yam, Gezari, Giomi, Goldstein, Golkhou, Goobar, Groom, Hacopians, Hale, Henning, Ho, Hover, Howell, Hung, Huppenkothen, Imel, Ip, Ivezić, Jackson, Jones, Juric, Kasliwal, Kaspi, Kaye, Kelley, Kowalski, Kramer, Kupfer, Landry, Laher, Lee, Lin, Lin, Lunnan, Giomi, Mahabal, Mao, Miller, Monkewitz, Murphy, Ngeow, Nordin, Nugent, Ofek, Patterson, Penprase, Porter, Rauch, Rebbapragada, Reiley, Rigault, Rodriguez, Roestel, Rusholme, Santen, Schulze, Shupe, Singer, Soumagnac, Stein, Surace, Sollerman, Szkody, Taddia, Terek, Van~Sistine, van Velzen, Vestrand, Walters, Ward, Ye, Yu, Yan, \& Zolkower}]{Bellm_2019}
Bellm, E.~C., Kulkarni, S.~R., Graham, M.~J., {et~al.} 2018, Publications of the Astronomical Society of the Pacific, 131, 018002, \dodoi{10.1088/1538-3873/aaecbe}

\bibitem[{{Bentz} {et~al.}(2010){Bentz}, {Walsh}, {Barth}, {Yoshii}, {Woo}, {Wang}, {Treu}, {Thornton}, {Street}, {Steele}, {Silverman}, {Serduke}, {Sakata}, {Minezaki}, {Malkan}, {Li}, {Lee}, {Hiner}, {Hidas}, {Greene}, {Gates}, {Ganeshalingam}, {Filippenko}, {Canalizo}, {Bennert}, \& {Baliber}}]{2010ApJ...716..993B}
{Bentz}, M.~C., {Walsh}, J.~L., {Barth}, A.~J., {et~al.} 2010, \apj, 716, 993, \dodoi{10.1088/0004-637X/716/2/993}

\bibitem[{Chambers {et~al.}(2019)Chambers, Magnier, Metcalfe, Flewelling, Huber, Waters, Denneau, Draper, Farrow, Finkbeiner, Holmberg, Koppenhoefer, Price, Rest, Saglia, Schlafly, Smartt, Sweeney, Wainscoat, Burgett, Chastel, Grav, Heasley, Hodapp, Jedicke, Kaiser, Kudritzki, Luppino, Lupton, Monet, Morgan, Onaka, Shiao, Stubbs, Tonry, White, Bañados, Bell, Bender, Bernard, Boegner, Boffi, Botticella, Calamida, Casertano, Chen, Chen, Cole, Deacon, Frenk, Fitzsimmons, Gezari, Gibbs, Goessl, Goggia, Gourgue, Goldman, Grant, Grebel, Hambly, Hasinger, Heavens, Heckman, Henderson, Henning, Holman, Hopp, Ip, Isani, Jackson, Keyes, Koekemoer, Kotak, Le, Liska, Long, Lucey, Liu, Martin, Masci, McLean, Mindel, Misra, Morganson, Murphy, Obaika, Narayan, Nieto-Santisteban, Norberg, Peacock, Pier, Postman, Primak, Rae, Rai, Riess, Riffeser, Rix, Röser, Russel, Rutz, Schilbach, Schultz, Scolnic, Strolger, Szalay, Seitz, Small, Smith, Soderblom, Taylor, Thomson, Taylor, Thakar, Thiel, Thilker, Unger, Urata, Valenti,
  Wagner, Walder, Walter, Watters, Werner, Wood-Vasey, \& Wyse}]{chambers2019panstarrs1surveys}
Chambers, K.~C., Magnier, E.~A., Metcalfe, N., {et~al.} 2019, The Pan-STARRS1 Surveys.
\newblock \doarXiv{1612.05560}

\bibitem[{Chanchaiworawit \& Sarajedini(2024)}]{Chanchaiworawit_2024}
Chanchaiworawit, K., \& Sarajedini, V. 2024, The Astrophysical Journal, 969, 131, \dodoi{10.3847/1538-4357/ad479a}

\bibitem[{{Charalampopoulos} {et~al.}(2022){Charalampopoulos}, {Leloudas}, {Malesani}, {Wevers}, {Arcavi}, {Nicholl}, {Pursiainen}, {Lawrence}, {Anderson}, {Benetti}, {Cannizzaro}, {Chen}, {Galbany}, {Gromadzki}, {Guti{\'e}rrez}, {Inserra}, {Jonker}, {M{\"u}ller-Bravo}, {Onori}, {Short}, {Sollerman}, \& {Young}}]{Charalampopoulos2022}
{Charalampopoulos}, P., {Leloudas}, G., {Malesani}, D.~B., {et~al.} 2022, \aap, 659, A34, \dodoi{10.1051/0004-6361/202142122}

\bibitem[{{Cohen} {et~al.}(1986){Cohen}, {Rudy}, {Puetter}, {Ake}, \& {Foltz}}]{1986ApJ...311..135C}
{Cohen}, R.~D., {Rudy}, R.~J., {Puetter}, R.~C., {Ake}, T.~B., \& {Foltz}, C.~B. 1986, \apj, 311, 135, \dodoi{10.1086/164758}

\bibitem[{{Dawson} {et~al.}(2013){Dawson}, {Schlegel}, {Ahn}, {Anderson}, {Aubourg}, {Bailey}, {Barkhouser}, {Bautista}, {Beifiori}, {Berlind}, {Bhardwaj}, {Bizyaev}, {Blake}, {Blanton}, {Blomqvist}, {Bolton}, {Borde}, {Bovy}, {Brandt}, {Brewington}, {Brinkmann}, {Brown}, {Brownstein}, {Bundy}, {Busca}, {Carithers}, {Carnero}, {Carr}, {Chen}, {Comparat}, {Connolly}, {Cope}, {Croft}, {Cuesta}, {da Costa}, {Davenport}, {Delubac}, {de Putter}, {Dhital}, {Ealet}, {Ebelke}, {Eisenstein}, {Escoffier}, {Fan}, {Filiz Ak}, {Finley}, {Font-Ribera}, {G{\'e}nova-Santos}, {Gunn}, {Guo}, {Haggard}, {Hall}, {Hamilton}, {Harris}, {Harris}, {Ho}, {Hogg}, {Holder}, {Honscheid}, {Huehnerhoff}, {Jordan}, {Jordan}, {Kauffmann}, {Kazin}, {Kirkby}, {Klaene}, {Kneib}, {Le Goff}, {Lee}, {Long}, {Loomis}, {Lundgren}, {Lupton}, {Maia}, {Makler}, {Malanushenko}, {Malanushenko}, {Mandelbaum}, {Manera}, {Maraston}, {Margala}, {Masters}, {McBride}, {McDonald}, {McGreer}, {McMahon}, {Mena}, {Miralda-Escud{\'e}}, {Montero-Dorta},
  {Montesano}, {Muna}, {Myers}, {Naugle}, {Nichol}, {Noterdaeme}, {Nuza}, {Olmstead}, {Oravetz}, {Oravetz}, {Owen}, {Padmanabhan}, {Palanque-Delabrouille}, {Pan}, {Parejko}, {P{\^a}ris}, {Percival}, {P{\'e}rez-Fournon}, {P{\'e}rez-R{\`a}fols}, {Petitjean}, {Pfaffenberger}, {Pforr}, {Pieri}, {Prada}, {Price-Whelan}, {Raddick}, {Rebolo}, {Rich}, {Richards}, {Rockosi}, {Roe}, {Ross}, {Ross}, {Rossi}, {Rubi{\~n}o-Martin}, {Samushia}, {S{\'a}nchez}, {Sayres}, {Schmidt}, {Schneider}, {Sc{\'o}ccola}, {Seo}, {Shelden}, {Sheldon}, {Shen}, {Shu}, {Slosar}, {Smee}, {Snedden}, {Stauffer}, {Steele}, {Strauss}, {Streblyanska}, {Suzuki}, {Swanson}, {Tal}, {Tanaka}, {Thomas}, {Tinker}, {Tojeiro}, {Tremonti}, {Vargas Maga{\~n}a}, {Verde}, {Viel}, {Wake}, {Watson}, {Weaver}, {Weinberg}, {Weiner}, {West}, {White}, {Wood-Vasey}, {Yeche}, {Zehavi}, {Zhao}, \& {Zheng}}]{2013AJ....145...10D}
{Dawson}, K.~S., {Schlegel}, D.~J., {Ahn}, C.~P., {et~al.} 2013, \aj, 145, 10, \dodoi{10.1088/0004-6256/145/1/10}

\bibitem[{Denney(2012)}]{Denney_2012}
Denney, K.~D. 2012, The Astrophysical Journal, 759, 44, \dodoi{10.1088/0004-637X/759/1/44}

\bibitem[{{Denney} {et~al.}(2014){Denney}, {De Rosa}, {Croxall}, {Gupta}, {Bentz}, {Fausnaugh}, {Grier}, {Martini}, {Mathur}, {Peterson}, {Pogge}, \& {Shappee}}]{2014ApJ...796..134D}
{Denney}, K.~D., {De Rosa}, G., {Croxall}, K., {et~al.} 2014, \apj, 796, 134, \dodoi{10.1088/0004-637X/796/2/134}

\bibitem[{{Ferland} {et~al.}(1990){Ferland}, {Korista}, \& {Peterson}}]{1990ApJ...363L..21F}
{Ferland}, G.~J., {Korista}, K.~T., \& {Peterson}, B.~M. 1990, \apjl, 363, L21, \dodoi{10.1086/185856}

\bibitem[{Flewelling {et~al.}(2020)Flewelling, Magnier, Chambers, Heasley, Holmberg, Huber, Sweeney, Waters, Calamida, Casertano, Chen, Farrow, Hasinger, Henderson, Long, Metcalfe, Narayan, Nieto-Santisteban, Norberg, Rest, Saglia, Szalay, Thakar, Tonry, Valenti, Werner, White, Denneau, Draper, Hodapp, Jedicke, Kaiser, Kudritzki, Price, Wainscoat, Chastel, McLean, Postman, \& Shiao}]{Flewelling_2020}
Flewelling, H.~A., Magnier, E.~A., Chambers, K.~C., {et~al.} 2020, The Astrophysical Journal Supplement Series, 251, 7, \dodoi{10.3847/1538-4365/abb82d}

\bibitem[{Gaskell \& Rojas~Lobos(2013)}]{10.1093/mnrasl/slt154}
Gaskell, C.~M., \& Rojas~Lobos, P.~A. 2013, Monthly Notices of the Royal Astronomical Society: Letters, 438, L36, \dodoi{10.1093/mnrasl/slt154}

\bibitem[{Green {et~al.}(2022)Green, Pulgarin-Duque, Anderson, MacLeod, Eracleous, Ruan, Runnoe, Graham, Roulston, Schneider, Ahlf, Bizyaev, Brownstein, del Casal, Dodd, Hoover, Matt, Merloni, Pan, Ramirez, Ridder, \& Moseley}]{Green_2022}
Green, P.~J., Pulgarin-Duque, L., Anderson, S.~F., {et~al.} 2022, The Astrophysical Journal, 933, 180, \dodoi{10.3847/1538-4357/ac743f}

\bibitem[{{Grier} {et~al.}(2012){Grier}, {Peterson}, {Pogge}, {Denney}, {Bentz}, {Martini}, {Sergeev}, {Kaspi}, {Zu}, {Kochanek}, {Shappee}, {Stanek}, {Araya Salvo}, {Beatty}, {Bird}, {Bord}, {Borman}, {Che}, {Chen}, {Cohen}, {Dietrich}, {Doroshenko}, {Efimov}, {Free}, {Ginsburg}, {Henderson}, {Horne}, {King}, {Mogren}, {Molina}, {Mosquera}, {Nazarov}, {Okhmat}, {Pejcha}, {Rafter}, {Shields}, {Skowron}, {Szczygiel}, {Valluri}, \& {van Saders}}]{2012ApJ...744L...4G}
{Grier}, C.~J., {Peterson}, B.~M., {Pogge}, R.~W., {et~al.} 2012, \apjl, 744, L4, \dodoi{10.1088/2041-8205/744/1/L4}

\bibitem[{{Guo} {et~al.}(2018){Guo}, {Shen}, \& {Wang}}]{2018ascl.soft09008G}
{Guo}, H., {Shen}, Y., \& {Wang}, S. 2018, {PyQSOFit: Python code to fit the spectrum of quasars}, Astrophysics Source Code Library, record ascl:1809.008

\bibitem[{Guo {et~al.}(2019)Guo, Sun, Liu, Wang, Kong, Wang, Sheng, \& He}]{Guo_2019}
Guo, H., Sun, M., Liu, X., {et~al.} 2019, The Astrophysical Journal Letters, 883, L44, \dodoi{10.3847/2041-8213/ab4138}

\bibitem[{{Guo} {et~al.}(2024){Guo}, {Zou}, {Fawcett}, {Canning}, {Juneau}, {Davis}, {Alexander}, {Jiang}, {Aguilar}, {Ahlen}, {Brooks}, {Claybaugh}, {de la Macorra}, {Doel}, {Fanning}, {Forero-Romero}, {Gontcho A Gontcho}, {Honscheid}, {Kisner}, {Kremin}, {Landriau}, {Meisner}, {Miquel}, {Moustakas}, {Nie}, {Pan}, {Poppett}, {Prada}, {Rezaie}, {Rossi}, {Siudek}, {Sanchez}, {Schubnell}, {Seo}, {Sui}, {Tarl{\'e}}, \& {Zhou}}]{2024ApJS..270...26G}
{Guo}, W.-J., {Zou}, H., {Fawcett}, V.~A., {et~al.} 2024, \apjs, 270, 26, \dodoi{10.3847/1538-4365/ad118a}

\bibitem[{{Guo} {et~al.}(2025){Guo}, {Pan}, {Siudek}, {Aguilar}, {Ahlen}, {Bianchi}, {Brooks}, {Claybaugh}, {Dawson}, {de la Macorra}, {Doel}, {Fanning}, {Forero-Romero}, {Gazta{\~n}aga}, {Gontcho A Gontcho}, {Honscheid}, {Kehoe}, {Kisner}, {Lambert}, {Landriau}, {Le Guillou}, {Manera}, {Meisner}, {Moustakas}, {Mu{\~n}oz-Guti{\'e}rrez}, {Myers}, {Nie}, {Palanque-Delabrouille}, {Poppett}, {Prada}, {Rezaie}, {Rossi}, {Sanchez}, {Schubnell}, {Seo}, {Silber}, {Sprayberry}, {Tarl{\'e}}, {Weaver}, {Zhou}, \& {Zou}}]{guo2024identificationlymanalphachanginglook}
{Guo}, W.-J., {Pan}, Z., {Siudek}, M., {et~al.} 2025, \apjl, 981, L8, \dodoi{10.3847/2041-8213/adb426}

\bibitem[{Guo {et~al.}(2025)Guo, Zou, Greenwell, Alexander, Fawcett, Pan, Siudek, Aguilar, Ahlen, Brooks, Claybaugh, Dawson, de~la Macorra, Doel, Font-Ribera, Gaztañaga, Gontcho A~Gontcho, Gutierrez, Kehoe, Kisner, Landriau, Le~Guillou, Manera, Meisner, Miquel, Moustakas, Prada, Rossi, Sanchez, Schubnell, Sprayberry, Sui, Tarlé, Weaver, Xiao, \& Zou}]{Guo_2025}
Guo, W.-J., Zou, H., Greenwell, C.~L., {et~al.} 2025, The Astrophysical Journal Supplement Series, 278, 28, \dodoi{10.3847/1538-4365/adc124}

\bibitem[{Guo~ {et~al.}(2020)Guo~, Shen, He, Wang, Liu, Wang, Sun, Yang, Kong, \& Sheng}]{Guo2020}
Guo~, H.~., Shen, Y., He, Z., {et~al.} 2020, The Astrophysical Journal, 888, 58, \dodoi{10.3847/1538-4357/ab5db0}

\bibitem[{{Hung} {et~al.}(2017){Hung}, {Gezari}, {Blagorodnova}, {Roth}, {Cenko}, {Kulkarni}, {Horesh}, {Arcavi}, {McCully}, {Yan}, {Lunnan}, {Fremling}, {Cao}, {Nugent}, \& {Wozniak}}]{2017ApJ...842...29H}
{Hung}, T., {Gezari}, S., {Blagorodnova}, N., {et~al.} 2017, \apj, 842, 29, \dodoi{10.3847/1538-4357/aa7337}

\bibitem[{{LaMassa} {et~al.}(2015){LaMassa}, {Cales}, {Moran}, {Myers}, {Richards}, {Eracleous}, {Heckman}, {Gallo}, \& {Urry}}]{2015ApJ...800..144L}
{LaMassa}, S.~M., {Cales}, S., {Moran}, E.~C., {et~al.} 2015, \apj, 800, 144, \dodoi{10.1088/0004-637X/800/2/144}

\bibitem[{LaMassa {et~al.}(2015)LaMassa, Cales, Moran, Myers, Richards, Eracleous, Heckman, Gallo, \& Urry}]{LaMassa_2015}
LaMassa, S.~M., Cales, S., Moran, E.~C., {et~al.} 2015, The Astrophysical Journal, 800, 144, \dodoi{10.1088/0004-637X/800/2/144}

\bibitem[{{Li} {et~al.}(2017){Li}, {Shen}, {Horne}, {Brandt}, {Greene}, {Grier}, {Ho}, {Kochanek}, {Schneider}, {Trump}, {Dawson}, {Pan}, {Bizyaev}, {Oravetz}, {Simmons}, \& {Malanushenko}}]{2017ApJ...846...79L}
{Li}, J., {Shen}, Y., {Horne}, K., {et~al.} 2017, \apj, 846, 79, \dodoi{10.3847/1538-4357/aa845d}

\bibitem[{{MacLeod} {et~al.}(2016){MacLeod}, {Ross}, {Lawrence}, {Goad}, {Horne}, {Burgett}, {Chambers}, {Flewelling}, {Hodapp}, {Kaiser}, {Magnier}, {Wainscoat}, \& {Waters}}]{2016MNRAS.457..389M}
{MacLeod}, C.~L., {Ross}, N.~P., {Lawrence}, A., {et~al.} 2016, \mnras, 457, 389, \dodoi{10.1093/mnras/stv2997}

\bibitem[{MacLeod {et~al.}(2019)MacLeod, Green, Anderson, Bruce, Eracleous, Graham, Homan, Lawrence, LeBleu, Ross, Ruan, Runnoe, Stern, Burgett, Chambers, Kaiser, Magnier, \& Metcalfe}]{MacLeod_2019}
MacLeod, C.~L., Green, P.~J., Anderson, S.~F., {et~al.} 2019, The Astrophysical Journal, 874, 8, \dodoi{10.3847/1538-4357/ab05e2}

\bibitem[{Magnier {et~al.}(2020)Magnier, Chambers, Flewelling, Hoblitt, Huber, Price, Sweeney, Waters, Denneau, Draper, Hodapp, Jedicke, Kaiser, Kudritzki, Metcalfe, Stubbs, \& Wainscoat}]{Magnier_2020}
Magnier, E.~A., Chambers, K.~C., Flewelling, H.~A., {et~al.} 2020, The Astrophysical Journal Supplement Series, 251, 3, \dodoi{10.3847/1538-4365/abb829}

\bibitem[{Margala {et~al.}(2016)Margala, Kirkby, Dawson, Bailey, Blanton, \& Schneider}]{Margala_2016}
Margala, D., Kirkby, D., Dawson, K., {et~al.} 2016, The Astrophysical Journal, 831, 157, \dodoi{10.3847/0004-637X/831/2/157}

\bibitem[{Masci {et~al.}(2018)Masci, Laher, Rusholme, Shupe, Groom, Surace, Jackson, Monkewitz, Beck, Flynn, Terek, Landry, Hacopians, Desai, Howell, Brooke, Imel, Wachter, Ye, Lin, Cenko, Cunningham, Rebbapragada, Bue, Miller, Mahabal, Bellm, Patterson, Jurić, Golkhou, Ofek, Walters, Graham, Kasliwal, Dekany, Kupfer, Burdge, Cannella, Barlow, Sistine, Giomi, Fremling, Blagorodnova, Levitan, Riddle, Smith, Helou, Prince, \& Kulkarni}]{Masci_2019}
Masci, F.~J., Laher, R.~R., Rusholme, B., {et~al.} 2018, Publications of the Astronomical Society of the Pacific, 131, 018003, \dodoi{10.1088/1538-3873/aae8ac}

\bibitem[{{Matt} {et~al.}(2003){Matt}, {Guainazzi}, \& {Maiolino}}]{2003MNRAS.342..422M}
{Matt}, G., {Guainazzi}, M., \& {Maiolino}, R. 2003, \mnras, 342, 422, \dodoi{10.1046/j.1365-8711.2003.06539.x}

\bibitem[{Miniutti {et~al.}(2013)Miniutti, Sanfrutos, Beuchert, Agís-González, Longinotti, Piconcelli, Krongold, Guainazzi, Bianchi, Matt, \& Jiménez-Bailón}]{Miniutti_2013}
Miniutti, G., Sanfrutos, M., Beuchert, T., {et~al.} 2013, Monthly Notices of the Royal Astronomical Society, 437, 1776–1790, \dodoi{10.1093/mnras/stt2005}

\bibitem[{{Osterbrock}(1977)}]{1977ApJ...215..733O}
{Osterbrock}, D.~E. 1977, \apj, 215, 733, \dodoi{10.1086/155407}

\bibitem[{{Osterbrock}(1981)}]{1981ApJ...249..462O}
---. 1981, \apj, 249, 462, \dodoi{10.1086/159306}

\bibitem[{{Osterbrock} \& {Koski}(1976)}]{1976MNRAS.176P..61O}
{Osterbrock}, D.~E., \& {Koski}, A.~T. 1976, \mnras, 176, 61P, \dodoi{10.1093/mnras/176.1.61P}

\bibitem[{Peterson {et~al.}(2000)Peterson, McHardy, Wilkes, Berlind, Bertram, Calkins, Collier, Huchra, Mathur, Papadakis, Peters, Pogge, Romano, Tokarz, Uttley, Vestergaard, \& Wagner}]{Peterson_2000}
Peterson, B.~M., McHardy, I.~M., Wilkes, B.~J., {et~al.} 2000, The Astrophysical Journal, 542, 161–174, \dodoi{10.1086/309518}

\bibitem[{Potts \& Villforth(2021)}]{Potts_2021}
Potts, B., \& Villforth, C. 2021, Astronomy \& Astrophysics, 650, A33, \dodoi{10.1051/0004-6361/202140597}

\bibitem[{{Rees}(1988)}]{1988Natur.333..523R}
{Rees}, M.~J. 1988, \nat, 333, 523, \dodoi{10.1038/333523a0}

\bibitem[{{Ren} {et~al.}(2024){Ren}, {Guo}, {Shen}, {Silverman}, {Burke}, {Wang}, \& {Wang}}]{2024ApJ...974..153R}
{Ren}, W., {Guo}, H., {Shen}, Y., {et~al.} 2024, \apj, 974, 153, \dodoi{10.3847/1538-4357/ad6e76}

\bibitem[{{Ren} {et~al.}(2022){Ren}, {Wang}, {Cai}, \& {Guo}}]{Ren2022}
{Ren}, W., {Wang}, J., {Cai}, Z., \& {Guo}, H. 2022, \apj, 925, 50, \dodoi{10.3847/1538-4357/ac3828}

\bibitem[{Ren {et~al.}(2024)Ren, Wang, Cai, \& Hu}]{Ren_2024}
Ren, W., Wang, J., Cai, Z., \& Hu, X. 2024, The Astrophysical Journal, 963, 7, \dodoi{10.3847/1538-4357/ad17cb}

\bibitem[{{Risaliti} {et~al.}(2002){Risaliti}, {Elvis}, \& {Nicastro}}]{2002ApJ...571..234R}
{Risaliti}, G., {Elvis}, M., \& {Nicastro}, F. 2002, \apj, 571, 234, \dodoi{10.1086/324146}

\bibitem[{{Ross} {et~al.}(2020){Ross}, {Graham}, {Calderone}, {Ford}, {McKernan}, \& {Stern}}]{2020MNRAS.498.2339R}
{Ross}, N.~P., {Graham}, M.~J., {Calderone}, G., {et~al.} 2020, \mnras, 498, 2339, \dodoi{10.1093/mnras/staa2415}

\bibitem[{Ruan {et~al.}(2016)Ruan, Anderson, Cales, Eracleous, Green, Morganson, Runnoe, Shen, Wilkinson, Blanton, Dwelly, Georgakakis, Greene, LaMassa, Merloni, \& Schneider}]{Ruan_2016}
Ruan, J.~J., Anderson, S.~F., Cales, S.~L., {et~al.} 2016, The Astrophysical Journal, 826, 188, \dodoi{10.3847/0004-637X/826/2/188}

\bibitem[{{Rumbaugh} {et~al.}(2018){Rumbaugh}, {Shen}, {Morganson}, {Liu}, {Banerji}, {McMahon}, {Abdalla}, {Benoit-L{\'e}vy}, {Bertin}, {Brooks}, {Buckley-Geer}, {Capozzi}, {Carnero Rosell}, {Carrasco Kind}, {Carretero}, {Cunha}, {D'Andrea}, {da Costa}, {DePoy}, {Desai}, {Doel}, {Frieman}, {Garc{\'\i}a-Bellido}, {Gruen}, {Gruendl}, {Gschwend}, {Gutierrez}, {Honscheid}, {James}, {Kuehn}, {Kuhlmann}, {Kuropatkin}, {Lima}, {Maia}, {Marshall}, {Martini}, {Menanteau}, {Plazas}, {Reil}, {Roodman}, {Sanchez}, {Scarpine}, {Schindler}, {Schubnell}, {Sheldon}, {Smith}, {Soares-Santos}, {Sobreira}, {Suchyta}, {Swanson}, {Walker}, {Wester}, \& {DES Collaboration}}]{Rumbaugh2018}
{Rumbaugh}, N., {Shen}, Y., {Morganson}, E., {et~al.} 2018, \apj, 854, 160, \dodoi{10.3847/1538-4357/aaa9b6}

\bibitem[{Sergeev(2020)}]{10.1093/mnras/staa1210}
Sergeev, S.~G. 2020, Monthly Notices of the Royal Astronomical Society, 495, 971, \dodoi{10.1093/mnras/staa1210}

\bibitem[{{Shen} {et~al.}(2011){Shen}, {Richards}, {Strauss}, {Hall}, {Schneider}, {Snedden}, {Bizyaev}, {Brewington}, {Malanushenko}, {Malanushenko}, {Oravetz}, {Pan}, \& {Simmons}}]{2011ApJS..194...45S}
{Shen}, Y., {Richards}, G.~T., {Strauss}, M.~A., {et~al.} 2011, \apjs, 194, 45, \dodoi{10.1088/0067-0049/194/2/45}

\bibitem[{{Shen} {et~al.}(2015){Shen}, {Brandt}, {Dawson}, {Hall}, {McGreer}, {Anderson}, {Chen}, {Denney}, {Eftekharzadeh}, {Fan}, {Gao}, {Green}, {Greene}, {Ho}, {Horne}, {Jiang}, {Kelly}, {Kinemuchi}, {Kochanek}, {P{\^a}ris}, {Peters}, {Peterson}, {Petitjean}, {Ponder}, {Richards}, {Schneider}, {Seth}, {Smith}, {Strauss}, {Tao}, {Trump}, {Wood-Vasey}, {Zu}, {Eisenstein}, {Pan}, {Bizyaev}, {Malanushenko}, {Malanushenko}, \& {Oravetz}}]{Shen2015}
{Shen}, Y., {Brandt}, W.~N., {Dawson}, K.~S., {et~al.} 2015, \apjs, 216, 4, \dodoi{10.1088/0067-0049/216/1/4}

\bibitem[{{Shen} {et~al.}(2019){Shen}, {Hall}, {Horne}, {Zhu}, {McGreer}, {Simm}, {Trump}, {Kinemuchi}, {Brandt}, {Green}, {Grier}, {Guo}, {Ho}, {Homayouni}, {Jiang}, {I-Hsiu Li}, {Morganson}, {Petitjean}, {Richards}, {Schneider}, {Starkey}, {Wang}, {Chambers}, {Kaiser}, {Kudritzki}, {Magnier}, \& {Waters}}]{2019ApJS..241...34S}
{Shen}, Y., {Hall}, P.~B., {Horne}, K., {et~al.} 2019, \apjs, 241, 34, \dodoi{10.3847/1538-4365/ab074f}

\bibitem[{Shen {et~al.}(2024)Shen, Grier, Horne, Stone, Li, Yang, Homayouni, Trump, Anderson, Brandt, Hall, Ho, Jiang, Petitjean, Schneider, Tao, Donnan, AlSayyad, Bershady, Blanton, Bizyaev, Bundy, Chen, Davis, Dawson, Fan, Greene, Groller, Guo, Ibarra-Medel, Jiang, Keenan, Kollmeier, Lejoly, Li, de~la Macorra, Moe, Nie, Rossi, Smith, Tee, Weijmans, Xu, Yue, Zhou, Zhou, \& Zou}]{shen2024sloandigitalskysurvey}
Shen, Y., Grier, C.~J., Horne, K., {et~al.} 2024, The Sloan Digital Sky Survey Reverberation Mapping Project: Key Results.
\newblock \doarXiv{2305.01014}

\bibitem[{{Sheng} {et~al.}(2017){Sheng}, {Wang}, {Jiang}, {Yang}, {Yan}, {Dou}, \& {Peng}}]{Sheng2017}
{Sheng}, Z., {Wang}, T., {Jiang}, N., {et~al.} 2017, \apjl, 846, L7, \dodoi{10.3847/2041-8213/aa85de}

\bibitem[{{STScI}(2022)}]{https://doi.org/10.17909/s0zg-jx37}
{STScI}. 2022, Pan-STARRS1 DR2 Catalog,  STScI/MAST, \dodoi{10.17909/S0ZG-JX37}

\bibitem[{{Sun} {et~al.}(2015){Sun}, {Trump}, {Shen}, {Brandt}, {Dawson}, {Denney}, {Hall}, {Ho}, {Horne}, {Jiang}, {Richards}, {Schneider}, {Bizyaev}, {Kinemuchi}, {Oravetz}, {Pan}, \& {Simmons}}]{2015ApJ...811...42S}
{Sun}, M., {Trump}, J.~R., {Shen}, Y., {et~al.} 2015, \apj, 811, 42, \dodoi{10.1088/0004-637X/811/1/42}

\bibitem[{van Velzen {et~al.}(2021)van Velzen, Gezari, Hammerstein, Roth, Frederick, Ward, Hung, Cenko, Stein, Perley, Taggart, Foley, Sollerman, Blagorodnova, Andreoni, Bellm, Brinnel, De, Dekany, Feeney, Fremling, Giomi, Golkhou, Graham, Ho, Kasliwal, Kilpatrick, Kulkarni, Kupfer, Laher, Mahabal, Masci, Miller, Nordin, Riddle, Rusholme, Santen, Sharma, Shupe, \& Soumagnac}]{Velzen_2021}
van Velzen, S., Gezari, S., Hammerstein, E., {et~al.} 2021, The Astrophysical Journal, 908, 4, \dodoi{10.3847/1538-4357/abc258}

\bibitem[{{Wang} {et~al.}(2018){Wang}, {Xu}, \& {Wei}}]{2018ApJ...858...49W}
{Wang}, J., {Xu}, D.~W., \& {Wei}, J.~Y. 2018, \apj, 858, 49, \dodoi{10.3847/1538-4357/aab88b}

\bibitem[{Wang {et~al.}(2020)Wang, Shen, Jiang, Grier, Horne, Homayouni, Peterson, Trump, Brandt, Hall, Ho, Li, Santisteban, Kinemuchi, McGreer, \& Schneider}]{Wang_2020}
Wang, S., Shen, Y., Jiang, L., {et~al.} 2020, The Astrophysical Journal, 903, 51, \dodoi{10.3847/1538-4357/abb36d}

\bibitem[{Wang {et~al.}(2024)Wang, Woo, Gallo, Guo, Son, Kong, Mandal, Cho, Kim, \& Shin}]{wang2024identifyingchanginglookagnsusing}
Wang, S., Woo, J.-H., Gallo, E., {et~al.} 2024, Identifying changing-look AGNs using variability characteristics.
\newblock \doarXiv{2402.18131}

\bibitem[{Waters {et~al.}(2020)Waters, Magnier, Price, Chambers, Burgett, Draper, Flewelling, Hodapp, Huber, Jedicke, Kaiser, Kudritzki, Lupton, Metcalfe, Rest, Sweeney, Tonry, Wainscoat, \& Wood-Vasey}]{Waters_2020}
Waters, C.~Z., Magnier, E.~A., Price, P.~A., {et~al.} 2020, The Astrophysical Journal Supplement Series, 251, 4, \dodoi{10.3847/1538-4365/abb82b}

\bibitem[{{Welsh} {et~al.}(2011){Welsh}, {Wheatley}, \& {Neil}}]{Welsh2011}
{Welsh}, B.~Y., {Wheatley}, J.~M., \& {Neil}, J.~D. 2011, \aap, 527, A15, \dodoi{10.1051/0004-6361/201015865}

\bibitem[{Winkler(1992)}]{10.1093/mnras/257.4.677}
Winkler, H. 1992, Monthly Notices of the Royal Astronomical Society, 257, 677, \dodoi{10.1093/mnras/257.4.677}

\bibitem[{{Yang} {et~al.}(2018){Yang}, {Wu}, {Fan}, {Jiang}, {McGreer}, {Shangguan}, {Yao}, {Wang}, {Joshi}, {Green}, {Wang}, {Feng}, {Fu}, {Yang}, \& {Liu}}]{2018ApJ...862..109Y}
{Yang}, Q., {Wu}, X.-B., {Fan}, X., {et~al.} 2018, \apj, 862, 109, \dodoi{10.3847/1538-4357/aaca3a}

\bibitem[{Yang {et~al.}(2020)Yang, Shen, Chen, Liu, Annis, Avila, Bertin, Brooks, Buckley-Geer, Carnero Rosell, Carrasco Kind, Carretero, da Costa, Desai, Thomas Diehl, Doel, Frieman, Garcia-Bellido, Gaztanaga, Gerdes, Gruen, Gruendl, Gschwend, Gutierrez, Hollowood, Honscheid, Hoyle, James, Krause, Kuehn, Lidman, Lima, Maia, Marshall, Martini, Menanteau, Miquel, Plazas Malagón, Sanchez, Scarpine, Schindler, Schubnell, Serrano, Sevilla, Smith, Soares-Santos, Sobreira, Suchyta, Swanson, Tarle, Vikram, \& Walker}]{10.1093/mnras/staa645}
Yang, Q., Shen, Y., Chen, Y.-C., {et~al.} 2020, Monthly Notices of the Royal Astronomical Society, 493, 5773, \dodoi{10.1093/mnras/staa645}

\bibitem[{Yip {et~al.}(2004{\natexlab{a}})Yip, Connolly, Vanden~Berk, Ma, Frieman, SubbaRao, Szalay, Richards, Hall, Schneider, Hopkins, Trump, \& Brinkmann}]{Yip_2004q}
Yip, C.~W., Connolly, A.~J., Vanden~Berk, D.~E., {et~al.} 2004{\natexlab{a}}, The Astronomical Journal, 128, 2603, \dodoi{10.1086/425626}

\bibitem[{Yip {et~al.}(2004{\natexlab{b}})Yip, Connolly, Szalay, Budavári, SubbaRao, Frieman, Nichol, Hopkins, York, Okamura, Brinkmann, Csabai, Thakar, Fukugita, \& Ivezić}]{Yip_2004g}
Yip, C.~W., Connolly, A.~J., Szalay, A.~S., {et~al.} 2004{\natexlab{b}}, The Astronomical Journal, 128, 585, \dodoi{10.1086/422429}

\bibitem[{Zeltyn {et~al.}(2024)Zeltyn, Trakhtenbrot, Eracleous, Yang, Green, Anderson, LaMassa, Runnoe, Assef, Bauer, Brandt, Davis, Frederick, Fries, Graham, Grogin, Guolo, Hernández-García, Koekemoer, Krumpe, Liu, Martínez-Aldama, Ricci, Schneider, Shen, Śniegowska, Temple, Trump, Xue, Brownstein, Dwelly, Morrison, Bizyaev, Pan, \& Kollmeier}]{Zeltyn_2024}
Zeltyn, G., Trakhtenbrot, B., Eracleous, M., {et~al.} 2024, The Astrophysical Journal, 966, 85, \dodoi{10.3847/1538-4357/ad2f30}

\bibitem[{{Zhu} {et~al.}(2016){Zhu}, {Wang}, {Cai}, \& {Sun}}]{Zhu2016}
{Zhu}, F.-F., {Wang}, J.-X., {Cai}, Z.-Y., \& {Sun}, Y.-H. 2016, \apj, 832, 75, \dodoi{10.3847/0004-637X/832/1/75}

\bibitem[{{ZTF Team}(2025)}]{https://doi.org/10.26131/irsa598}
{ZTF Team}. 2025, ZTF Lightcurves,  IPAC, \dodoi{10.26131/IRSA598}

\bibitem[{Zuo {et~al.}(2012)Zuo, Wu, Liu, \& Jiao}]{Zuo_2012}
Zuo, W., Wu, X.-B., Liu, Y.-Q., \& Jiao, C.-L. 2012, The Astrophysical Journal, 758, 104, \dodoi{10.1088/0004-637X/758/2/104}

\end{thebibliography}
\bibliographystyle{aasjournal}

\end{document}